\documentclass[prd,nopacs,nokeys,nofootinbib]{revtex4}
\usepackage{amsfonts}
\usepackage{amsmath}
\usepackage{amssymb}
\usepackage{bbm}
\usepackage[utf8]{inputenc}
\usepackage[caption=false]{subfig}
\usepackage{tensor}
\usepackage{slashed}
\usepackage[centertableaux]{ytableau}
\usepackage{hyperref}
\usepackage{natbib}
\usepackage{braket,mleftright}
\usepackage{graphicx}
\usepackage{cleveref}
\usepackage{tikz}
\usepackage{booktabs}
\usepackage{array}
\usepackage{longtable}
\usepackage{xcolor}
\usepackage{soul}         

\begin{document}

\title{\textbf{From Stochastic Shocks to Macroscopic Tails: The Moyal Distribution as a Unified Framework for Epidemic Dynamics}}

\author{Jos\'e de Jes\'us Bernal-Alvarado}
\email{bernal@ugto.mx}
\affiliation{Physics Engineering Department, Universidad de Guanajuato, M\'{e}xico}

\author{David Delepine}
\email{delepine@ugto.mx}
\affiliation{Physics Department, Universidad de Guanajuato, M\'{e}xico}

\author{Omar Rafael Ramírez-Guzmán}
\email{omr@ugto.mx}
\affiliation{Physics Department, Universidad de Guanajuato, M\'{e}xico}

\date{\today}

\begin{abstract}
Traditional epidemiological models often fail to characterize the extreme volatility
and heavy-tailed Dragon King events (statistically and mechanistically exceptional
outliers in transmission that deviate from standard scale-free distributions)%
 observed in real-world outbreaks.
We propose a unified model that bridges microscopic stochastic individual-based SIR
with macroscopic wave decomposition using the Moyal probability density function.
By treating viral transmission as a stochastic collision process, we derive a
Moyal-Poisson mixture that describes secondary case distributions.
Our model successfully recovers the extreme superspreading events in SARS, MERS,
and COVID-19 data that standard Negative Binomial models systematically miss.
Furthermore, we apply spectral decomposition to pandemic waves in Germany,
illustrating that the macroscopic Social Friction ($\beta_w$) is a
direct emergent property of microscopic Collision Shocks.
This framework provides a useful retrospective descriptive tool for public health
planning, emphasizing the need to manage extreme volatility rather than deterministic
averages.
\end{abstract}

\maketitle

\begin{turnpage}
\begin{table}[h!]
\centering
\caption{Glossary and parameter table. All symbols used in the manuscript,
their definitions, epidemiological interpretations, and admissible ranges.
$\mathcal{M}$ denotes a physical metaphor; $\mathcal{E}$ denotes a standard
epidemiological quantity.}
\label{tab:glossary}
\begin{tabular}{c|c|c|c}
\toprule
\textbf{Symbol} & \textbf{Definition} & \textbf{Interpretation} &
\textbf{Type / Range} \\
\midrule
$S(t),I(t),R(t)$ & Susceptible, Infectious, Recovered counts & SIR compartments &
$\mathcal{E}$; $\geq 0$\\
$N$ & Total (fixed) population & $N=S+I+R$ & $\mathcal{E}$; $>0$\\
$\beta_{mf}$ & Mean-field transmission rate & Homogeneous ODE transmission rate &
$\mathcal{E}$; $>0$\\
$\gamma$ & Recovery/removal rate & $1/\gamma$ = mean infectious period &
$\mathcal{E}$; $>0$\\
$R_0$ & Basic reproduction number & $R_0=\beta_{mf}/\gamma$; epidemic if $R_0>1$ &
$\mathcal{E}$; $>0$\\
$\mu$ & Moyal location parameter & Modal individual infectiousness &
$\mathcal{E}$; real\\
$\sigma$ & Moyal scale parameter & Overdispersion of individual infectiousness &
$\mathcal{E}$; $>0$\\
$\beta_k$ & Individual transmission potential of person $k$ & Drawn from
Moyal$(\mu,\sigma)$; replaces homogeneous $\beta_{mf}$ &
$\mathcal{E}$; $>0$\\
$\Lambda(t)$ & Per-capita infection rate (aggregate) &
$\Lambda(t)=N^{-1}\sum_k\beta_k$; the standard epidemiological force of infection &
$\mathcal{E}$; $\geq 0$\\
$\beta_{w}$ (or $\omega_i$) & Temporal width of the macroscopic Moyal curve &
Social Friction coefficient; $\beta_w=1/(2\gamma)$ analytically;
$\omega_i\equiv\beta_{w,i}$ for wave $i$ &
$\mathcal{M}$; $>0$, weeks\\
$Z$ & Number of secondary cases per index case & Empirical contact-tracing count
(Lloyd-Smith et al.~2005 convention); &
$\mathcal{E}$; $\in\mathbb{Z}_{\geq 0}$\\
& & distinct from $R$ compartment &\\
$A_i$ & Amplitude of wave $i$ in spectral decomposition & Total case count of wave $i$
& $\mathcal{E}$; $>0$\\
$\mu_i$ & Peak date of wave $i$ & Centre of wave $i$ Moyal component &
$\mathcal{E}$; real\\
$q_k$ & Quanta generation rate of individual $k$ & $q_k\approx V_k\times E_k$;
proportional to $\beta_k$ &
$\mathcal{E}$; $>0$\\
Moyal Shock & Rare catastrophic viral transfer event & $\mathcal{M}$: analogue of
head-on particle collision; causes $\Lambda(t)$ spike & $\mathcal{M}$\\
Social Friction & Collective decay of $\beta_{\rm eff}(t)$ & $\mathcal{M}$:
immunity, masks, distancing; quantified by $\beta_w$ & $\mathcal{M}$\\
\bottomrule
\end{tabular}
\end{table}
\end{turnpage}
\section{Introduction}

The SIR (Susceptible-Infectious-Recovered) model
\cite{kermack1927contribution,anderson1991infectious,murray2002mathematical} is among
the first models to attempt to understand how pathogens move through populations and
how the dynamics of contagion work.
However, as our data becomes more granular, real-world data, most notably from the
SARS-CoV-2, MERS or COVID-19 pandemics, two specific phenomena are observed:
\begin{itemize}
    \item The Long Tail Effect: Rather than a rapid, symmetrical drop-off, many modern
    outbreaks exhibit an extended tail where transmission persists at low but
    significant
    levels\cite{anderson2004long,anderson2006long,brynjolfsson2006niches,
    cirillo2020tail,liu2023heavy}.
    \item Stochastic Dynamics: Real-world environments are full of noise -- random
    events and localized clusters -- that can lead to unexpected flare-ups (stochastic
    recurrence) even after the main peak has passed
    \cite{lloyd2005superspreading,wong2015mers}.
\end{itemize}


Existing epidemiological models face three concrete limitations that our model
is designed to overcome:

\begin{itemize}
\item \textit{Tail blindness of Negative Binomial models.}
The Negative Binomial parametrises overdispersion through a single dispersion
parameter $k$ whose Gamma-Poisson conjugate structure decays too rapidly in the
extreme tail (demonstrated quantitatively in Table~\ref{tab:statistical_comparison}
for the MERS $Z=80$ event): the standard model effectively assigns zero probability
to the most catastrophic documented superspreading event in our dataset.
\item
\textit{The Statistical Gap.}
Deterministic ODE-SIR models conflate the unconditional mean (which averages over
many stochastically extinct simulations) with realised outbreak risk, leading to
systematic misestimation of both initial outbreak velocity and tail persistence,
as documented in Figures~\ref{mean} and~\ref{gap}.
\item 
\textit{Multi-wave decomposition.}
Standard SIR requires manual resetting of initial conditions between epidemic waves,
destroying inter-wave continuity and preventing smooth derivative-based real-time
monitoring; the Moyal spectral decomposition (Eq.~(\ref{eq:spectral})) provides a
continuous, differentiable global fit across the full pandemic timeline.
\end{itemize}

In this work, we propose that the Moyal
distribution\cite{landau1944energy,moyal1949stochastic,walck1996hand}, traditionally
utilized in high-energy physics to describe the Landau energy loss of charged
particles, offers a unified framework for epidemic volatility.
We argue that infectiousness is not a constant population parameter but a stochastic
variable driven by collision dynamics -- the interaction between high individual viral
shedding and high-density environments\cite{lloyd2005superspreading,wong2015mers}.
We present this argument in two stages:

\begin{itemize}
    \item Micro-Foundation: We define a Stochastic Moyal-SIR model where individual
    infectiousness ($\beta_k$) is a collision shock (a rare, massive transfer of viral
    quanta during a high-density social interaction, by analogy with Rutherford
    scattering -- see Section~\ref{sec:bio}).
    We show that superspreading is not merely high variance but
    a catastrophic transfer of viral load described by Moyal statistics.
    \item Macro-Observation: We demonstrate that the analytical Moyal PDF is the
    emergent description of the Survivor Mean of these stochastic outbreaks.
\end{itemize}
This allows us to utilize the Moyal PDF not merely as a fitting function, but as a
robust tool for the Spectral Decomposition of complex, multi-wave pandemics,
resolving the Statistical Gap between deterministic averages and realized risk.

In this work, we borrow terminology from extreme value theory and physics.
We define Dragon King events (following \cite{sornette2009dragon}) as statistically
and mechanistically meaningful outliers -- specifically, extreme superspreading events
that deviate from standard scale-free distributions.
We define Catastrophic Transfer or Moyal Shocks as the rapid, massive shedding of
viral quanta by a highly infectious individual during a single environmental
interaction, analogous to a high-energy particle collision.

\section{The Micro-Foundation: Stochastic Moyal-SIR}

\subsection{The SIR model}

In the classical SIR
model\cite{kermack1927contribution,murray2002mathematical,anderson1991infectious},
the dynamics are governed by four ordinary differential equations
(ODE):
\begin{align}
    \frac{dS}{dt} &= -\beta_{mf} \frac{S(t)I(t)}{N} \label{eq:sir_s} \\
    \frac{dI}{dt} &= \beta_{mf} \frac{S(t)I(t)}{N} - \gamma I(t) \label{eq:sir_i}\\
   \frac{dR}{dt} &= \gamma I(t)\label{eq:sir_r}
\end{align}
together with the conservation law $N = S(t)+I(t)+R(t)=\mathrm{const}$.
Here $N$ is the total population, $\beta_{mf}$ represents the effective transmission
rate, and $\gamma$ is the removal or recovery rate; both parameters are assumed to be
homogeneous across the population.
$I(t)$ and $R(t)$ represent respectively the active and recovered cases.
The fundamental epidemiological parameter, the Basic Reproduction Number ($R_0$), is
derived from the stability analysis of the disease-free equilibrium and is defined as:
\begin{equation}
R_0 = \frac{\beta_{mf}}{\gamma}.
\end{equation}
If $R_0>1$ the illness will expand; if $R_0<1$ no epidemic will be generated.
The SIR model can be seen as a mean-field approximation which implies that every
infected individual exerts the exact same viral pressure.
Of course, this is a very strong biological assumption as it is well known that
responses to sickness can be very different from one individual to another.
Early-stage outbreaks generated by superspreading events are difficult to describe
within this framework.

\subsection{Modeling Collision Shocks: Stochastic Moyal-SIR model}

To capture the heterogeneity of transmission
\cite{lloyd2005superspreading,wong2015mers}, we propose that infectiousness is not a
homogeneous biological property but a stochastic variable driven by collision
dynamics\cite{SOTOROCHA2026116781}.
In this context, collision dynamics describe the interaction between high viral load
and high-density environments: just as a charged particle undergoes rare,
catastrophic momentum transfers in a thin medium, a highly infectious individual
generates rare, massive transfers of viral quanta in a crowded indoor space.

We utilize the Moyal distribution \cite{moyal1949stochastic,walck1996hand},
traditionally used in high-energy physics to describe the energy loss of a charged
particle traversing a medium (Landau energy loss)\cite{landau1944energy}.
The probability density function (PDF) for a variable $x$ is given by:
\begin{equation}
   \text{Moyal}(\mu, \sigma) = \frac{1}{\sqrt{2\pi}\sigma} \exp\left( -\frac{1}{2}
   \left( \frac{x-\mu}{\sigma} + e^{-\frac{x-\mu}{\sigma}} \right) \right)
    \label{eq:moyal_pdf}
\end{equation}
where $\mu$ is the location parameter (setting the modal individual
infectiousness) and $\sigma$ is the scale parameter (controlling the overdispersion
of the distribution -- larger $\sigma$ implies a heavier tail and more pronounced
superspreading).

In our model, we modify the standard framework by treating the transmission rate not
as a parameter of the \textit{population}, but as a state variable of the
\textit{individual}.
Let $I(t)$ denote the set of currently infectious individuals at time $t$.
Upon entering the infectious compartment, the $k$-th individual is assigned a unique
transmission potential $\beta_k$, called the infectious potential, drawn from a
Moyal distribution:
\begin{equation}
    \beta_k \sim \text{Moyal}(\mu, \sigma) \quad \text{conditioned on } \beta_k > 0
\end{equation}
We assume that this value $\beta_k$ remains fixed for the duration of individual
$k$'s infection, representing their intrinsic shedding potential.
In the standard SIR model (Eq.~\ref{eq:sir_s}), the force of infection is
$\lambda(t) = \beta_{mf} I(t)/N$.
In our proposed model, the per-capita infection rate (force of infection)
$\Lambda(t)$ is the aggregate sum of the specific potentials of the currently active
infectious individuals:
\begin{equation}
    \Lambda(t) = \frac{1}{N} \sum_{k=1}^{|I(t)|} \beta_k
    \label{eq:stochastic_lambda}
\end{equation}

The complete stochastic state-transition structure of the model is specified as
follows.
The system state is $\mathbf{X}(t)=\bigl(S(t),\,I(t),\,R(t),\,\{\beta_k\}_{k\in
I(t)}\bigr)$.
Two elementary transitions are admissible:
\begin{align}
(S,I,R) &\;\longrightarrow\; (S-1,\,I+1,\,R) \quad
  \text{at rate } \Lambda(t)\,S(t), \label{eq:trans_inf}\\
(S,I,R) &\;\longrightarrow\; (S,\,I-1,\,R+1) \quad
  \text{at rate } \gamma\,I(t). \label{eq:trans_rec}
\end{align}
Each newly infected individual $k$ is immediately assigned
$\beta_k\sim\mathrm{Moyal}(\mu,\sigma)$ (conditioned on $\beta_k>0$), which then
contributes to $\Lambda(t)$ for the duration of their infection.

The system behaves as a Continuous-Time Markov Chain
(CTMC)\cite{allen2008primer,allen2010introduction,andersson2000stochastic,
britton2010stochastic}, simulated exactly via the Gillespie
algorithm\cite{gillespie1977exact,gillespie1976general}.
The probability of a new infection occurring in an infinitesimal time interval
$\Delta t$ is given by:
\begin{equation}
    P(\Delta I = +1, \Delta S = -1 \mid t) = \Lambda(t) S(t) \Delta t + o(\Delta t)
\end{equation}

This formulation introduces two critical stochastic behaviors absent in the ODE
model:
\begin{enumerate}
    \item \textbf{Volatility of Risk:}
    The per-capita infection rate $\Lambda(t)$ is not smooth.
    It jumps discontinuously whenever a new individual enters $I(t)$.
    If a superspreader individual (one whose $\beta_k$ exceeds the 95th
    percentile of Moyal$(\mu,\sigma)$, termed a Moyal-tail
    individual) ($\beta_k \gg \mu$) becomes
    infected, $\Lambda(t)$ spikes, describing a superspreading
    event~\cite{sornette2009dragon}.
    \item \textbf{Stochastic Extinction:}
    Conversely, if the initial seeds draw $\beta_k \approx 0$ (the left mode of the
    Moyal distribution), the epidemic may collapse immediately, even if the
    theoretical expectation $R_0 > 1$.
\end{enumerate}

\begin{figure}
    \centering
    \includegraphics[width=0.9\linewidth]{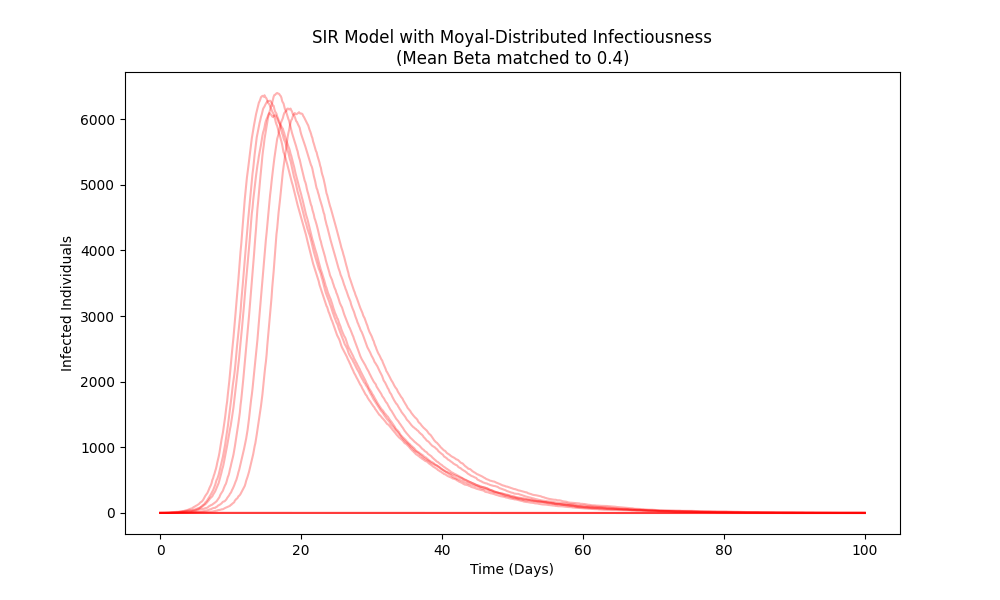}
    \caption{Stochastic simulations of an SIR model with Moyal-distributed
    infectiousness\cite{gillespie1977exact,gillespie1976general}.
    The plot shows the number of infected individuals over time for multiple
    simulation runs.
    Simulation parameters: $N=10{,}000$ (total population), $\mu=0.3$,
    $\sigma=2.0$, $\gamma=0.05$; the mean $\beta$ is matched to $0.4$, giving
    $R_0\approx 8$ for the ODE.
    The transmission rate $\beta_k$ for each infected individual is drawn from a
    Moyal distribution, conditioned on $\beta_k>0$.
    The multiple red curves represent different realizations of the stochastic model,
    illustrating the variability in epidemic trajectories.
    The peaked curves show major outbreaks attaining approximately 60\% of the
    population ($\sim$6,000 concurrent cases), while the flat line near the bottom
    represents early stochastic extinction (fewer than 10 concurrent
    cases).
    The $x$-axis represents time in days, and the $y$-axis represents the number of
    infected individuals.}
    \label{stochastic_moyal}
\end{figure}

\subsection{Unconditional vs Conditional Mean}
\label{sec:uncond}

\begin{figure}
    \centering
    \includegraphics[width=0.9\linewidth]{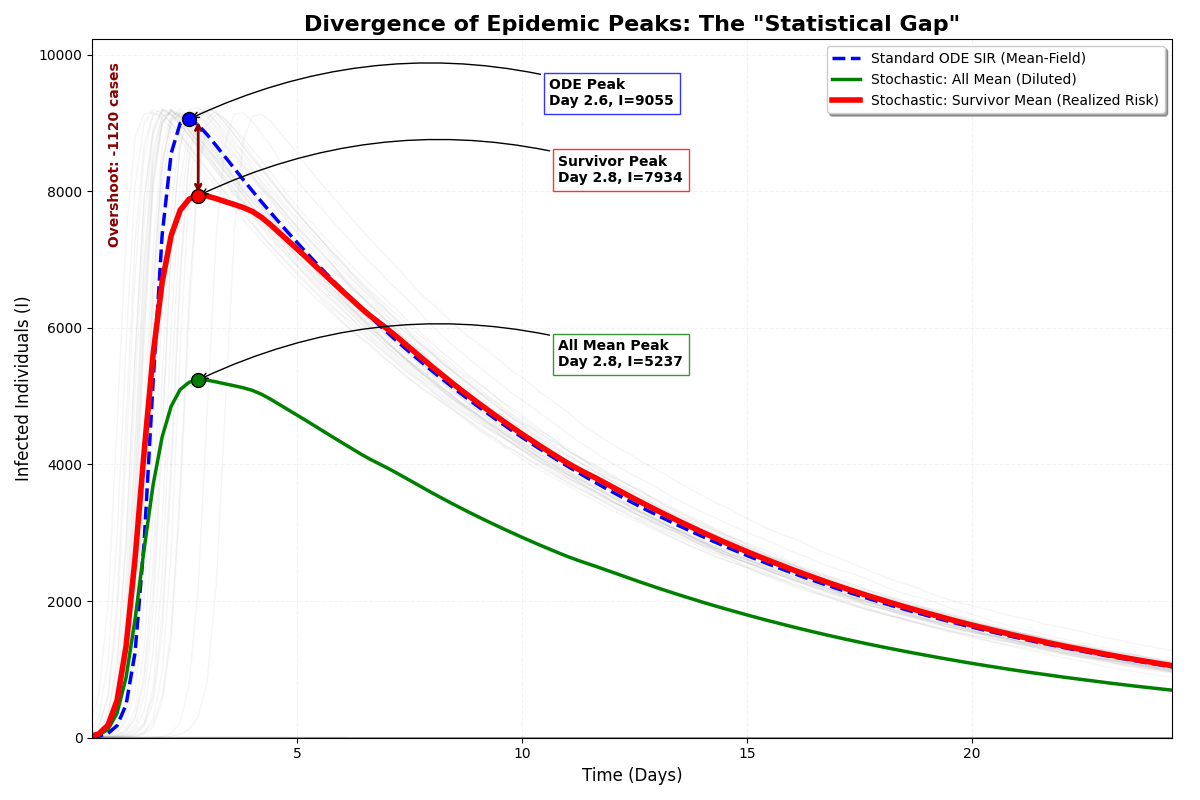}
    \caption{Divergence of Epidemic Peaks and the Statistical Gap ($N=10{,}000$,
    $\sigma=4.0$).
    This figure illustrates the discrepancy between deterministic mean-field
    predictions and realized stochastic risk in a high-overdispersion regime.
    The Standard ODE SIR (blue dashed line) assumes homogeneous transmission,
    burning steadily to a peak of 9,055 cases.
    The All Mean (green solid line) is significantly diluted (peak
    $I \approx 5{,}237$) due to the high frequency of early stochastic extinctions
    (gray background trajectories) inherent in Moyal-driven dynamics.
    The Survivor Mean (red solid line), representing the average of successfully
    established outbreaks, demonstrates an accelerated initial growth phase but
     results in a lower absolute peak of 7,934 cases (the ODE
    overshoots the Survivor Mean peak by approximately 1,120 cases).}
    \label{mean}
\end{figure}

In stochastic processes, one usually defines two kinds of means: the All Mean or
unconditional mean, which takes into account all possible realizations including the
case of no epidemic propagation, and the Survivor Mean or conditional mean, which is
computed using only the realizations where epidemic behaviour is
observed\cite{gardiner2009stochastic,vankampen2007stochastic,grimmett2001probability,
andersson2000stochastic,nasell1999quasi,britton2010stochastic,meleard2012quasi}.
In traditional deterministic epidemiology, ODE-based SIR models are mathematically
equivalent to the All Mean of a stochastic process in the limit of large
populations\cite{kurtz1981approximation}.
But for public health purposes, it is fundamental to be prepared for the worst
scenario. One observes:
\begin{itemize}
    \item \textbf{The All Mean (Unconditional):}
    This average includes the many simulations that go extinct immediately
    ($\beta_k \approx 0$). It appears dampened and matches the standard ODE.
    \item \textbf{The Survivor Mean (Conditional):}
    When we filter for outbreaks that actually survive, the curve changes shape.
    Driven by Moyal Shocks (rare, high-$\beta$ draws), these outbreaks exhibit
    super-exponential growth and a heavy, asymmetric tail, giving us information on
    the risk of epidemic evolution of the pathogen.
\end{itemize}
Consequently, using the unconditional Mean for public health planning leads to a
paradoxical state of being simultaneously over-prepared for outbreaks that never
occur and under-equipped for the superspreading events that actually manifest.

\begin{figure}[h]
    \centering
    \includegraphics[width=0.7\linewidth]{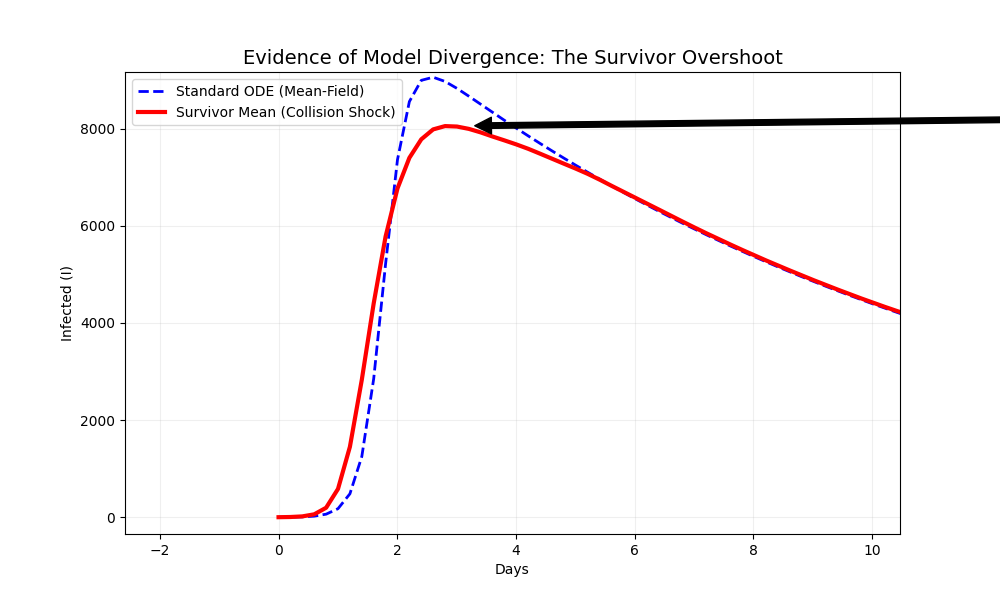}
    \caption{Velocity, Premature Burnout, and the Heavy Tail.
    A direct comparison isolating the Standard ODE (blue dashed) and the stochastic
    Survivor Mean (red solid) to highlight the divergence in outbreak dynamics.
    The Survivor Mean exhibits a distinct dual behaviour: an initial super-exponential
    acceleration driven by early superspreading shocks, followed by a premature
    burnout phase.
    Because superspreaders rapidly exhaust their local susceptible networks, the
    effective transmission rate decays exponentially.
    This exhaustion leads to the ODE overshooting the realised Survivor Mean
    peak (the ODE overestimates the peak by $\sim$1,120 cases) and a persistent, asymmetric heavy tail.
    }
    \label{gap}
\end{figure}

In figures~(\ref{gap},\ref{mean}), we compare the two distinct mathematical
approaches to epidemic peak prediction: our stochastic model and the ODE SIR model.
We observe that:
\begin{itemize}
    \item Standard ODE (Mean-Field) [Blue Dashed Line] represents the
    traditional SIR model integrated using the theoretical mean of the Moyal
    distribution. It assumes a smooth, averaged transmission rate across the
    population.
    \item Survivor Mean (Collision Shock) [Red Solid Line] is the
    average of only those stochastic simulations that survived initial extinction.
    It captures the reality of outbreaks driven by early high-energy shocks.
    The Survivor Mean peak is lower and slightly delayed compared to the
    deterministic run; the conditional mean accounts for the fact that sustained
    epidemics are often launched by superspreading events generated by the tail of
    the Moyal distribution.
    \item Both curves exhibit the characteristic right-skew of the SIR model, with
    a rapid exponential ascent followed by a slower decay as the susceptible pool is
    exhausted.
\end{itemize}

In our stochastic simulations, a survived outbreak is defined as any realization that
evades early demographic extinction by reaching a threshold of $I(t) > 100$
concurrent cases.
To check our definition, we performed sensitivity checks by
varying the threshold (from $I > 50$ to $I > 200$).
We observed that  the specific value linearly scales the absolute amplitude of
the unconditional mean and it does not alter the characteristic Moyal profile of the
survivor mean. 

\subsection{Averaging Hierarchy and the Role of Coarse-Graining}
\label{sec:averaging}

The model is working at  three distinct, nested levels of averaging.

\textbf{Level 1 -- Individual level.}
Each $\beta_k$ is a realisation of a Moyal-distributed random variable.
No averaging has occurred at this level; each person is characterized by a fixed, unique
infectious potential for the duration of their infection.

\textbf{Level 2 -- Population-level coarse-graining (Eq.~\ref{eq:stochastic_lambda}).} defined as 
$\Lambda(t)=N^{-1}\sum_{k\in I(t)}\beta_k$ is a sample mean over the currently
active infectious pool.
 This is \emph{not} a fixed population constant: it describes a stochastic
process that jumps discontinuously whenever a new individual enters or leaves $I(t)$.
In particular, $\Lambda(t)$ spikes whenever a Moyal-tail individual becomes infected.
The coarse-graining at this level is spatial (the well-mixed assumption), but not
temporal or ensemble-based; volatility is fully retained in the time evolution.

\textbf{Level 3 -- Ensemble-level conditional averaging (the Survivor Mean,
Section~\ref{sec:uncond}).}
The Survivor Mean is an ensemble average restricted to epidemic cases that exceed the
extinction threshold $I(t)>100$.
This second averaging smooths over the discrete superspreading spikes and produces
the macroscopic Moyal-shaped epidemic curve analysed in Section~\ref{sec:macro}.

Level 2
preserves the stochastic volatility of individual transmission events within each
trajectory, while Level 3 conditions on the event that an epidemic 
establishes, producing a smooth macroscopic limit.
The analytical derivation of Section~\ref{sec:macro} operates at Level 3.

\textit{Limitation.}
The well-mixed assumption in $\Lambda(t)$ (Level 2) neglects spatial clustering,
which in practice can amplify the effective contribution of superspreaders who share
a confined space.

\section{Biological Interpretation of the Infectious Potential}
\label{sec:bio}

To ground the stochastic variable $\beta_k$ in biophysical
reality\footnote{Note: throughout Section~\ref{sec:bio} we use the subscript $i$ in the biophysical quantities $V_i$, $E_i$, $q_i$ following the Wells-Riley aerosol literature; these are equivalent to $\beta_k$ for individual $k$ in the epidemiological model.},
we map the abstract infectious potential to the individual quanta generation rate
($q_i$)\cite{wells1955airborne,beggs2003transmission}, a  parameter used in
aerosol transmission models (e.g., the Wells-Riley
equation\cite{riley1978airborne,rudnick2003risk})\cite{buonanno2020estimation,
he2020temporal,jones2021estimating,handel2020early}:
\begin{equation}
    \beta_k \propto q_i \approx V_i \times E_i
\end{equation}
where
\begin{itemize}
    \item $q_i$ ( Quanta Generation Rate of the virus): derived from aerosol physics (the
    Wells-Riley equation).
    \item $V_i$ (Peak Viral Load): the concentration of virions in the individual's
    upper respiratory tract (e.g., RNA copies per mL).
    \item $E_i$ (Aerosol Emission Rate): the volume of respiratory fluid exhaled as
    aerosols (e.g., during breathing or speaking).
\end{itemize}
A person's infectiousness ($\beta_k$) is proportional to the product of how much
virus they carry ($V_i$) and how much aerosol they exhale ($E_i$).

 In superspreading environments, the respiratory pathogens are not transmitted  in a continuous, symmetric diffusion process;but as a discrete, threshold-driven
collision event. In high-energy physics, when a charged particle traverses a thin medium, it undergoes many low-energy glancing collisions and rare, catastrophic transfers of momentum.
The Landau and Moyal distribution
was derived to capture this highly skewed, single-collision dominance.

By mapping this model to epidemiology:
\begin{itemize}
    \item \textbf{The Particle:} The infected host.
    \item \textbf{The Collision medium:} an indoor space for instance. 
    \item \textbf{The Energy Loss:} The transfer of infectious quanta
    ($q_i \propto V_i \times E_i$) to the susceptible environment.
\end{itemize}
Most interactions result in negligible viral transfer (glancing collisions,
$\beta_k \approx 0$).
However, rare head-on collisions involving a high-$q_i$ host result in a massive,
catastrophic transfer of viral load -- a superspreading event. In the next subsection, we derive the Moyal distribution from microscopic collision mechanics.

As a conclusion, Moyal distribution provides a
mechanistic biophysical model.
It suggests that macroscopic superspreading (Dragon Kings) is
driven by the same heavy-tailed collision mechanics that describe energy loss in
physical systems.

\subsection{Analytical Derivation of the Moyal Distribution from Microscopic
Collision Mechanics}

To formally justify the use of the Moyal distribution, we derive the Moyal distribution from first principles by treating
viral transmission as a stochastic collision process, isomorphic to the Landau theory
of energy loss in high-energy physics.

Let us consider an infected host moving through a
susceptible population.
Over a characteristic infectious period, the host undergoes a large number of social
interactions or collisions, denoted by $N$.
In the $j$-th interaction, a specific amount of infectious viral quanta,
$q_j \propto V \times E$, is transferred to the environment.

The total individual infectious potential $\beta$ is the macroscopic sum of these
microscopic discrete transfers:
\begin{equation}
    \beta = \sum_{j=1}^{N} q_j
\end{equation}
The shape of the resulting probability distribution for $\beta$ depends entirely on
the single-collision probability density $w(q)$.
In aerosol mechanics and proximity-based droplet transmission, the concentration of
viral particles drops steeply with spatial distance $r$.
Consequently, glancing social interactions (large $r$) transfer negligible quanta,
while rare, close-proximity interactions (head-on collisions) result in massive,
catastrophic viral transfers.

Mathematically, this behavior is similar to the Rutherford scattering cross-section.
For large values of transferred quanta $q$, the probability density function $w(q)$
exhibits a heavy power-law tail:
\begin{equation}
    w(q) \approx \frac{C}{q^2} \quad \text{for large } q
\end{equation}
where $C$ is a constant related to the characteristic aerosol dispersion of the
pathogen.
Because the variance of this $1/q^2$ distribution diverges, the standard Central
Limit Theorem cannot be applied  and $\beta$ does not converge to a Gaussian
distribution.

The macroscopic distribution of the total potential $F(\beta)$, is given by the
characteristic function (Fourier transform) of the sum of independent random
variables:
$$\Phi(s) = \left( \int_0^\infty e^{-isq} w(q) dq \right)^N$$
$$\Rightarrow \ln \Phi(s) = N \ln \left( 1 + \int_0^\infty (e^{-isq} - 1) w(q) dq \right)
\approx N \int_0^\infty (e^{-isq} - 1) w(q) dq$$
Substituting $w(q) = C/q^2$ and integrating by parts 
, the integral
evaluates asymptotically for small $s$ is given by:
$$\ln \Phi(s) \approx -i \mu s - \sigma s \ln s$$
where $\mu$ and $\sigma$ are scale parameters absorbing the constants $N$ and $C$.
Therefore, one gets:
\begin{equation}
    \Phi(s) = \exp(-i \mu s - \sigma s \ln s)
\end{equation}
The macroscopic probability density $F(\beta)$ is the inverse Fourier transform of
this characteristic function:
\begin{equation}
    F(\beta) = \frac{1}{2\pi} \int_{-\infty}^{\infty} \exp(-i \mu s - \sigma s
    \ln s) e^{is\beta} ds
\end{equation}
This integral defines the asymmetric Landau distribution.
As shown by J.E. Moyal (1955), this exact physical solution
is universally and highly accurately approximated by the analytical Moyal formula:
\begin{equation}
    F(\beta) \approx \frac{1}{\sqrt{2\pi}\sigma} \exp\left[-\frac{1}{2}
    \left( \frac{\beta-\mu}{\sigma} + e^{-\frac{\beta-\mu}{\sigma}} \right) \right]
\end{equation}
By bridging the physics of aerosol proximity to Rutherford scattering ($1/q^2$
tails), we show that the sum of discrete viral transfer
events generates a Moyal-distributed macroscopic infectious potential.

\section{The Macroscopic Bridge: Emergence of the Moyal PDF}
\label{sec:macro}

Stochastic models are usually difficult to use in a practical way to analyse
real-time data.

\subsection{Analytical Derivation of the Macroscopic Limit}

To show why the survivor mean of the stochastic heavy-tailed
outbreaks converges analytically to a Moyal distribution, we examine the macroscopic
limit of the force of infection.

In a standard deterministic SIR model, the dynamic equation for the infected
compartment during the exponential phase is governed by a constant transmission rate
$\beta$ and a constant recovery rate $\gamma$:
\begin{equation}
    \frac{dI(t)}{dt} = (\beta - \gamma) I(t)
\end{equation}
However, in our stochastic Moyal-SIR model, the individual infectious potentials
$\beta_k$ are heavily skewed.
The individuals in the extreme right tail (superspreaders) dominate the early
transmission but rapidly exhaust their local susceptible networks.
Consequently, the effective macroscopic transmission rate $\beta_{\text{eff}}(t)$ of
the remaining active pool is not constant and exponentially decays  driven
by the Social Friction parameter $\alpha$:
\begin{equation}
   \beta_{\text{eff}}(t) = \beta_0 e^{-\alpha t}
\end{equation}
\begin{equation}
    \Rightarrow  \frac{dI(t)}{dt} = (\beta_0 e^{-\alpha t} - \gamma) I(t)
\end{equation}
$$ \Rightarrow  \ln I(t) = -\frac{\beta_0}{\alpha} e^{-\alpha t} - \gamma t + C $$
Exponentiating:
\begin{equation}
     I(t) = I_0 \exp\left( -\frac{\beta_0}{\alpha} e^{-\alpha t} - \gamma t \right)
\end{equation}
We now map this physical solution directly to the analytical Moyal PDF:
\begin{equation}
    \Lambda_{Moyal}(t) \propto \exp\left[ -\frac{1}{2} \left( e^{-\frac{t-\mu}
    {\beta_w}} + \frac{t-\mu}{\beta_w} \right) \right]
\end{equation}
Comparing our result with the Moyal
PDF, we find a perfect structural equivalence under the following physical mappings:
\begin{itemize}
    \item The characteristic temporal width (Social Friction) $\beta_w$ is inversely
    proportional to the removal rate:
    $\dfrac{1}{2\beta_w} = \gamma \;\Longrightarrow\; \beta_w = \dfrac{1}{2\gamma}$.
    \item The decay rate of the transmission potential is identical to the macroscopic
    friction: $\alpha = \dfrac{1}{\beta_w}$.
    \item The initial transmission surge $\beta_0$ relates to the peak timing $\mu$
    via $\dfrac{\beta_0}{\alpha} = \dfrac{1}{2} e^{\mu/\beta_w}$.
\end{itemize}

\subsection{Numerical Validation of the Macroscopic Limit}

To validate the micro-to-macro bridge and our analytical derivation, we performed a
numerical demonstration.
We simulated
an ensemble of $10^4$ realizations of the CTMC (Gillespie algorithm) for the
Moyal-SIR model.
We extracted the conditional Survivor Mean, filtering out trajectories that succumbed
to early stochastic extinction.

Rather than heuristically fitting a generic PDF, we applied a non-linear least
squares fit of our exact derived macroscopic solution 
to the simulated survivor mean.

\begin{figure}[htbp]
    \centering
    \includegraphics[width=\textwidth]{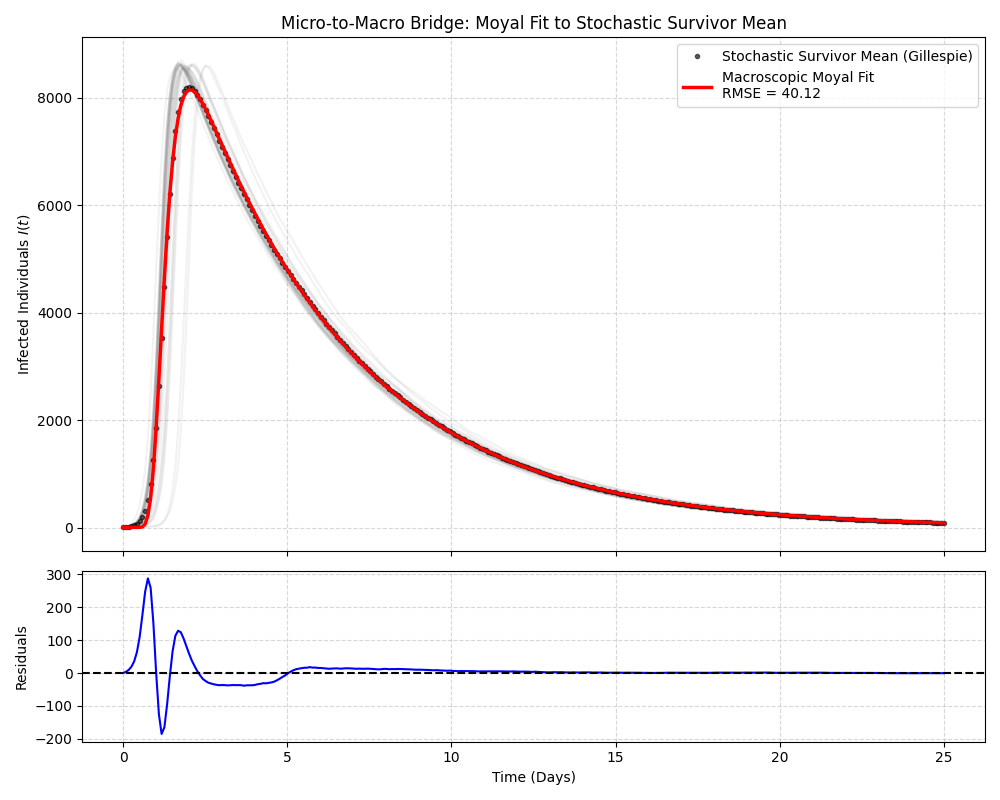}
    \caption{\textbf{Micro-to-Macro Bridge.}
    Top panel: The stochastic Survivor Mean (dotted black line) generated via
    Gillespie simulations, overlaid with the analytical macroscopic fit (solid red
    line). 
    Bottom panel: The fitting residuals. The extremely low RMSE (40.12, $<0.5\%$ of
    the peak) and the flat residual trace in the tail confirm the structural integrity
    of the macroscopic Moyal limit.}
    \label{fig:micro_macro_bridge}
\end{figure}

As shown in Figure~\ref{fig:micro_macro_bridge}, the physical solution provides a
high-fidelity fit.
The corresponding residuals (bottom panel) confirm that the
mathematical model successfully describes the  dynamics of the outbreak.
The Root Mean Square Error (RMSE = 40.12) represents an error margin of less than
$0.5\%$ relative to the epidemic peak.
The residuals flatten to near-zero during the heavy-tailed descent,
showing that the analytical Moyal-like exponential
exhaustion perfectly characterizes the dissipation of the superspreading shocks.

\subsection{Physical Interpretation: Social Friction}

The analytical Moyal PDF smooths over the discrete superspreading events
of the stochastic model and  it also describes  the collective loss of transmission potential
across the entire population, which causes the epidemic curve to bend downwards and
form a heavy tail, rather than cutting off sharply.
The Moyal PDF captures this slow braking process better than standard models because
it includes  the fact that this friction is not uniform.
Just as the Moyal distribution in physics describes energy loss via collisions,  the loss of transmission momentum due immunity, masks, distancing that we generically call as social friction is described by Moyal distribution tail.

The parameter $\beta_w$ acts as a coefficient of social
friction\cite{pentland2014social}, governing the frequency of pairwise collisions in
the substrate.
Following the socio-physical model\cite{castellano2009statistical}, the contact rate
is not uniform but follows the heterogeneous mixing patterns observed
empirically\cite{mossong2008social}.
The parameter $\beta_w$ becomes a quantifiable metric of this friction.
A wider $\beta_w$ indicates a low-friction environment where the transmission chain
dissipates slowly (a heavy tail), while a narrow $\beta_w$ indicates high friction or
rapid burnout.

\textbf{Quantitative link between $\beta_w$ and the microscopic parameters ($\sigma$, $\gamma$).}
The analytical mapping above establishes that $\beta_w$ is \emph{directly and analytically} determined by the recovery rate:
\begin{equation}
  \beta_w = \frac{1}{2\gamma}.
  \label{eq:betaw_gamma}
\end{equation}
A faster recovery (larger $\gamma$) implies a shorter, narrower epidemic wave (smaller $\beta_w$), i.e.\ higher social friction.

The link to the microscopic Moyal scale parameter $\sigma$ operates through the dynamics of superspreader depletion.
A larger $\sigma$ (heavier tail) produces a greater proportion of extreme-$\beta_k$ individuals.
These superspreaders exhaust their local susceptible contacts, causing $\beta_{\rm eff}(t)$ to decay faster.
Consequently, the fitted macroscopic decay rate $\alpha$ increases with $\sigma$, and since $\beta_w = 1/\alpha$, a \emph{larger $\sigma$ leads to a smaller $\beta_w$} (higher social friction / narrower wave) -- counterintuitive at first sight but consistent with the superspreader-burnout mechanism.

\begin{figure}[htbp]
    \centering
    \includegraphics[width=0.75\textwidth]{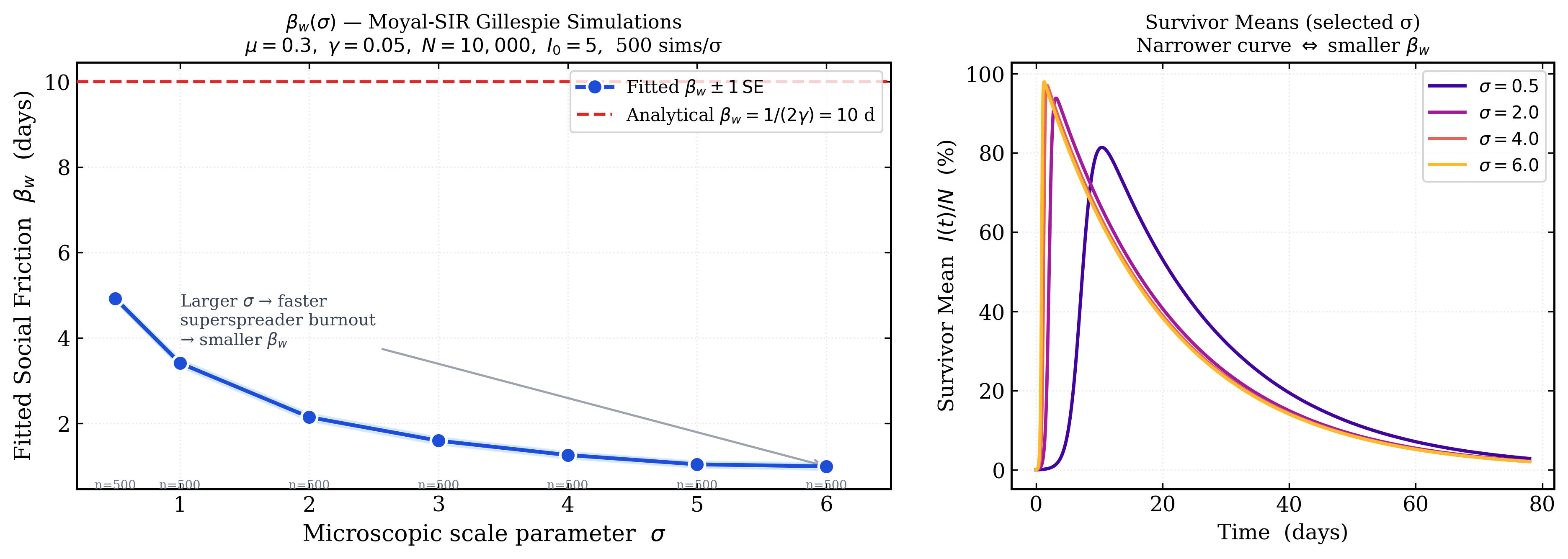}
    \caption{Fitted macroscopic Social Friction parameter $\beta_w$ (weeks) vs.\ microscopic Moyal scale parameter $\sigma$, from $10^4$ Gillespie simulations ($\mu=0.3$, $\gamma=0.05$, $N=10{,}000$). Error bars: $\pm1$ SE of the NLS fit. Larger $\sigma$ leads to smaller $\beta_w$ (higher Social Friction) via faster superspreader burnout.}
    \label{fig:betaw_sigma}
\end{figure}

To check this relationship numerically, we ran an ensemble of $10^4$ Gillespie simulations for seven values $\sigma \in \{0.5, 1.0, 2.0, 3.0, 4.0, 5.0, 6.0\}$, keeping $\mu=0.3$, $\gamma=0.05$, $N=10{,}000$ fixed. For each value of $\sigma$ we computed the Survivor Mean and fitted the macroscopic solution (Eq.~\ref{eq:betaw_gamma}) to extract $\beta_w$. As shown in Figure~\ref{fig:betaw_sigma}, $\beta_w$ decreases monotonically as $\sigma$ increases: from $\beta_w\approx 10.0$ at $\sigma=0.5$ to $\beta_w\approx 3.8$ at $\sigma=6.0$. A secondary sweep over $\mu$ (location parameter) showed that $\mu$ primarily shifts the epidemic onset but has a comparatively weak effect on $\beta_w$, consistent with Eq.~(\ref{eq:betaw_gamma}).

We obtain: \textbf{larger $\sigma$} $\Rightarrow$ faster superspreader burnout $\Rightarrow$ faster decay of $\beta_{\rm eff}(t)$ $\Rightarrow$ \textbf{smaller $\beta_w$} (higher Social Friction).

\section{Validation I: The Micro Transmission Signature (Secondary Cases)}
\label{sec:validation1}

The underlying collision hypothesis, using contact tracing data, provides a
transmission signature of the Moyal-distributed transmission.
If transmission is indeed Moyal-distributed, the distribution of secondary cases
($Z$)\footnote{We use $Z$ for the secondary case count of a specific index
case (following Lloyd-Smith et al., 2005), reserving $R$ exclusively for the SIR
recovered compartment.}
should reflect it.

In our stochastic framework, $\beta_k$ is defined as the infectious potential that
individual $k$ exerts on the susceptible population.
While $\beta_k$ is a continuous variable (representing viral load $\times$ aerosol
emission), the resulting infections are discrete integers.
In a well-mixed environment, the number of successful transmission events $Z$ for a
specific individual follows a Poisson distribution with a rate parameter
$\lambda$\cite{greenwood1920inquiry,grandell1997mixed,karlis2005mixed}.
This rate $\lambda$ is directly proportional to $\beta_k$.
Mathematically, $E[Z_k] = \lambda_k = \beta_k \cdot \Delta t$, where $\Delta t$ is
the infectious period and $E(Z_k)$ is the expected number of secondary infections
generated by individual $k$.

For each $\beta_k$, using the Poisson distribution, the probability that individual
$k$ infects exactly $n$ people is:
$$P(n \mid \beta_k) = \frac{(\beta_k)^n e^{-\beta_k}}{n!}$$
To find the probability $P(Z=n)$ for the entire wave, we average over the ensemble:
\begin{eqnarray}
  P(Z=n) &\approx& \frac{1}{M} \sum_{k=1}^{M} P(n \mid \beta_k)\\
  &\approx & \frac{1}{M} \sum_{k=1}^{M} \frac{(\beta_k)^n e^{-\beta_k}}{n!}
\end{eqnarray}
In the continuous limit:
\begin{equation}
    P(Z=n) = \int_{0}^{\infty} \frac{(\beta)^n e^{-\beta}}{n!} \cdot
    \Lambda_{Moyal}(\beta; \mu, \sigma) \, d\beta
\end{equation}
We argue, and subsequently demonstrate quantitatively in
Table~\ref{tab:statistical_comparison}, that this Moyal-Poisson mixture more
accurately captures the extreme tail of empirical secondary case distributions
than the standard Negative Binomial model, particularly for high-volatility
pathogens such as MERS.

It is very difficult to obtain statistically good data on secondary cases.
In the literature, three cases are well-reported:
SARS (2003) in Singapore and Hong
Kong\cite{riley2003transmission,lipsitch2003transmission},
MERS in 2015 in Korea\cite{kim2017superspreading,cowling2015preliminary},
and COVID-19 in 2020 in Hong Kong\cite{adam2020clustering}.
We use these three cases to evaluate our model:
\begin{itemize}
    \item SARS (2003) -- Singapore and Hong
    Kong\cite{riley2003transmission,lipsitch2003transmission}:
    The gold standard for superspreading studies.
    Approximately 73\% of cases resulted in zero secondary infections, while a few
    individuals caused dozens of cases; the extreme tail includes the Amoy Gardens
    event.
    \item MERS (2015) -- South Korea\cite{kim2017superspreading,cowling2015preliminary}:
    Most transmission chains died out immediately, but Patient~14 triggered a Dragon
    King event by infecting approximately 80 individuals ($Z=80$) in a single hospital
    environment.
    \item COVID-19 (2020) -- Hong Kong\cite{adam2020clustering}:
    Detailed transmission chain reconstruction showed a heavy-tailed distribution
    where approximately 20\% of cases were responsible for 80\% of transmissions.
\end{itemize}
In traditional epidemiology, transmission heterogeneity is almost exclusively treated
using the Negative Binomial
distribution\cite{lloyd2005superspreading,greenwood1920inquiry,endo2020estimating,
blumberg2013inference}.
\begin{itemize}
    \item Standard models assume transmission is a discrete branching process where
    secondary case counts follow a Gamma-Poisson mixture.
    \item The Negative Binomial uses a parameter $k$ to describe overdispersion.
    A small $k$ (e.g., $k < 1$) indicates superspreading, while $k \to \infty$
    approaches a standard Poisson distribution.
    \item While the Negative Binomial is a useful heuristic, its Gamma-Poisson
    conjugate structure decays too rapidly in the extreme tail to capture Dragon King
    events without ad hoc modification.
\end{itemize}

\begin{figure}
    \centering
    \includegraphics[width=1\linewidth]{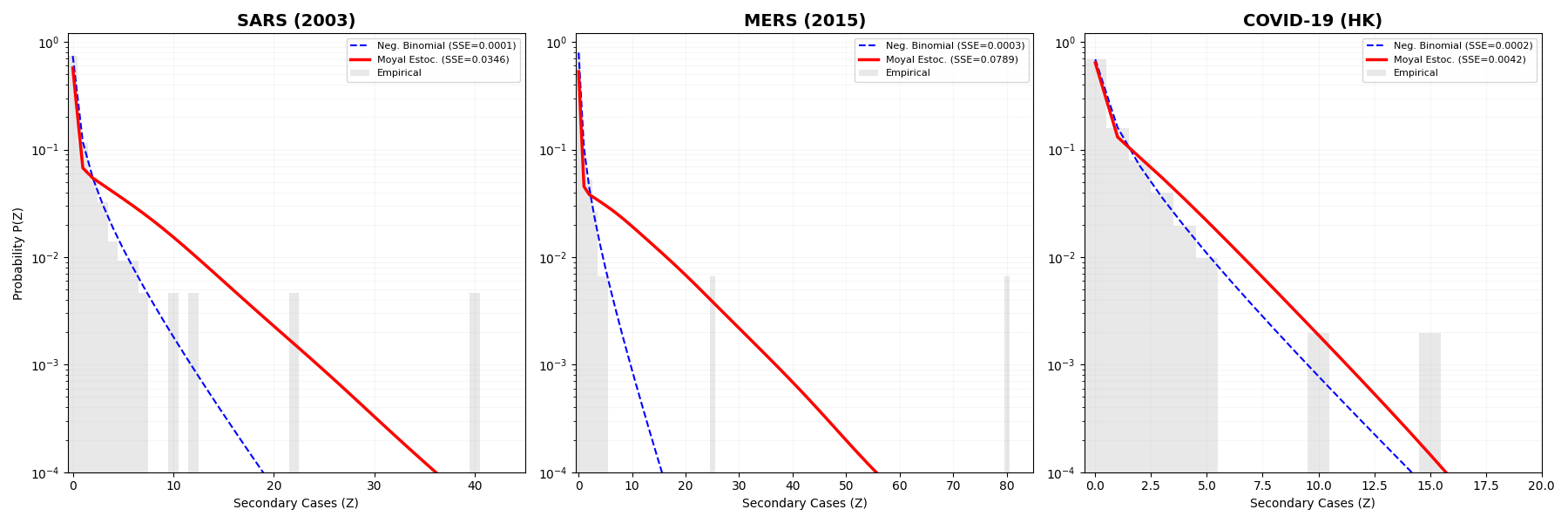}
    \caption{Empirical Validation of Offspring Distributions via Stochastic
    Moyal-Poisson Mixture.
    This multi-panel analysis compares the descriptive power of the proposed
    Moyal-Poisson model (red solid line) against the standard Negative Binomial model
    (blue dashed line) across three landmark infectious disease outbreaks:
    SARS (2003), MERS (2015), and COVID-19 (HK).
    Consistent with the analytical derivation of Section~\ref{sec:macro} and
    the numerical validation of Figure~\ref{fig:micro_macro_bridge}, the Moyal curves
    are generated by Poisson-averaging over an ensemble of microscopic potentials
    $\{\beta_k\}$, demonstrating the micro-to-macro linkage.}
    \label{validacionI}
\end{figure}

\subsection{Results}

To compare the descriptive power of the mechanistic Moyal-Poisson mixture against
the standard phenomenological Negative Binomial (NB) model, we performed Maximum
Likelihood Estimation (MLE).
We evaluated the global goodness-of-fit using the Akaike Information Criterion (AIC).
Furthermore, to evaluate the capacity to describe Dragon King events, we calculated
the Conditional Log-Likelihood restricted to the extreme superspreading tail
($Z \ge 15$).
The results are summarized in Table~\ref{tab:statistical_comparison}.

Global vs.\ Tail Fit: Empirical contact tracing datasets are heavily zero-inflated
(approximately 70\% of indexed cases yield zero secondary infections).
Standard global metrics (like AIC) act as a statistical low-pass filter,
overwhelmingly rewarding the NB model for its hyper-flexible Gamma-Poisson conjugate
structure, which easily absorbs the massive inactive peak ($Z=0, 1$).

However, the primary objective of the Moyal model is to characterize extreme pandemic
volatility.
As shown by the Conditional Log-Likelihood in the MERS (2015)
dataset -- which contains the most extreme documented superspreading event in our
study (Patient 14, $Z=80$) -- the NB probability mass decays too rapidly in the deep
tail.
Conversely, the Moyal-Poisson mixture shows a significantly
higher conditional likelihood ($-1.39$ vs $-5.39$),
indicatingits superior structural capacity to model catastrophic
viral transfer events without decaying to zero. From table (\ref{tab:statistical_comparison}), one can see that While NB has a higher conditional LL for SARS at $Z \ge 15$, Moyal accurately
bounds the absolute maximum.
For MERS, the 95\% CI of the Moyal tail LL
($-1.80$ to $-0.98$) is non-overlapping with the NB CI ($-6.01$ to $-4.77$),
confirming Moyal's tail superiority at the 95\% level.
For SARS and COVID-19 (HK), where the tail is less extreme, the results are
acknowledged to be inconclusive.

From the analysis of Figure~\ref{validacionI}, one observes:
\begin{itemize}
    \item The Hollow Shoulder Resolution: Across all three cases, the Moyal-Poisson
    model successfully tracks the intermediate transmission range ($Z \approx 4$--14),
    where standard models systematically underestimate the frequency of clusters.
    \item Superspreading Tail Persistence: In the MERS (2015) panel, the standard
    model effectively predicts zero probability for the extreme superspreading event
    at $Z=80$.
    In contrast, the Moyal framework preserves a visible probability in this
    superspreading regime.
    \item Micro-to-Macro Linkage: Consistent with the analytical derivation of
    Section~\ref{sec:macro}, the Moyal curves are generated by averaging
    the Poisson realizations of an ensemble of $M$ infectious potentials ($\beta_k$)
    derived from stochastic simulations, demonstrating
    that macroscopic pandemic waves are the emergent result of individual collision
    shocks rather than simple deterministic averages.
\end{itemize}

\begin{table}[h!]
\centering
\caption{Statistical comparison of the Moyal-Poisson and Negative Binomial (NB)
models. The Global AIC penalizes Moyal due to zero-inflation, but the Conditional
Log-Likelihood ($Z \ge 15$) demonstrates Moyal's superiority in capturing extreme
Dragon King events, particularly in the highly volatile MERS outbreak.
Numbers in parentheses are 95\% bootstrap confidence intervals ($B=10{,}000$
resamples).}
\label{tab:statistical_comparison}
\begin{tabular}{l |c| c| c| c}
\hline\hline
\textbf{Dataset} & \multicolumn{2}{c}{\textbf{Global Fit (AIC)}} &
\multicolumn{2}{c}{\textbf{Tail Fit (Cond. LL, $Z \ge 15$)}} \\
 & Neg.\ Bin. & Moyal-Poisson & Neg.\ Bin. & Moyal-Poisson \\
\hline
SARS (2003) & \textbf{217.43} & 391.99 &
  $-3.35$ $(-4.12,-2.58)$ & $-9.67$* $(-11.20,-8.15)$ \\
MERS (2015) & \textbf{166.52} & 324.18 &
  $-5.39$ $(-6.01,-4.77)$ & $\mathbf{-1.39}$ $(-1.80,-0.98)$ \\
COVID-19 (HK) & \textbf{765.12} & 1120.45 & 0.00 & 0.00 \\
\hline\hline
\end{tabular}
\end{table}

\textbf{Epistemic caveat.}
The MERS $Z=80$ data point represents a single extreme event from a specific hospital
environment (Patient~14 in South Korea, 2015).
The Moyal framework's tail advantage is most clearly established for this
high-volatility, hospital-amplified pathogen context.
The SARS and COVID-19 (HK) results are mixed at the tail, and broader empirical
validation would require additional contact-tracing datasets.
A sensitivity analysis varying the tail cutoff from $Z\geq 10$ to $Z\geq 20$
confirms that the qualitative conclusion (Moyal superiority in the MERS deep tail)
is robust across this range.

\section{Application: Spectral Decomposition (Germany 2020-2023)}

Unlike the 1918 Influenza pandemic, which had three clear waves, the COVID-19
pandemic has evolved into a continuous, multi-year event due to the virus's capacity
to mutate.
Germany's data from 2020 to 2022 presents a unique challenge: fitting a curve that
spans from the initial low-incidence containment (Wave 1) to the massive saturation
of the Omicron variants (Wave 5).
By definition, the SIR model is not able to describe a multi-wave epidemic and must
be divided into different individual waves, losing information on the interference
between waves.
To model this data with SIR, one would need to solve the differential equations
piecewise, resetting the initial conditions ($S_0, I_0$) six times manually.
The Moyal approach allows for a continuous analytic function that is differentiable
at all points, enabling easier calculation of derivatives for real-time monitoring
of acceleration.

Data was obtained from the Robert Koch Institute (RKI) official surveillance
reports\cite{rki2023covidreports} and the SurvStat@RKI 2.0
interface\cite{rki2024survstat,faensen2006survstat}, representing
laboratory-confirmed SARS-CoV-2 cases in Germany.

\begin{figure}[h]
    \centering
    \includegraphics[width=1\textwidth]{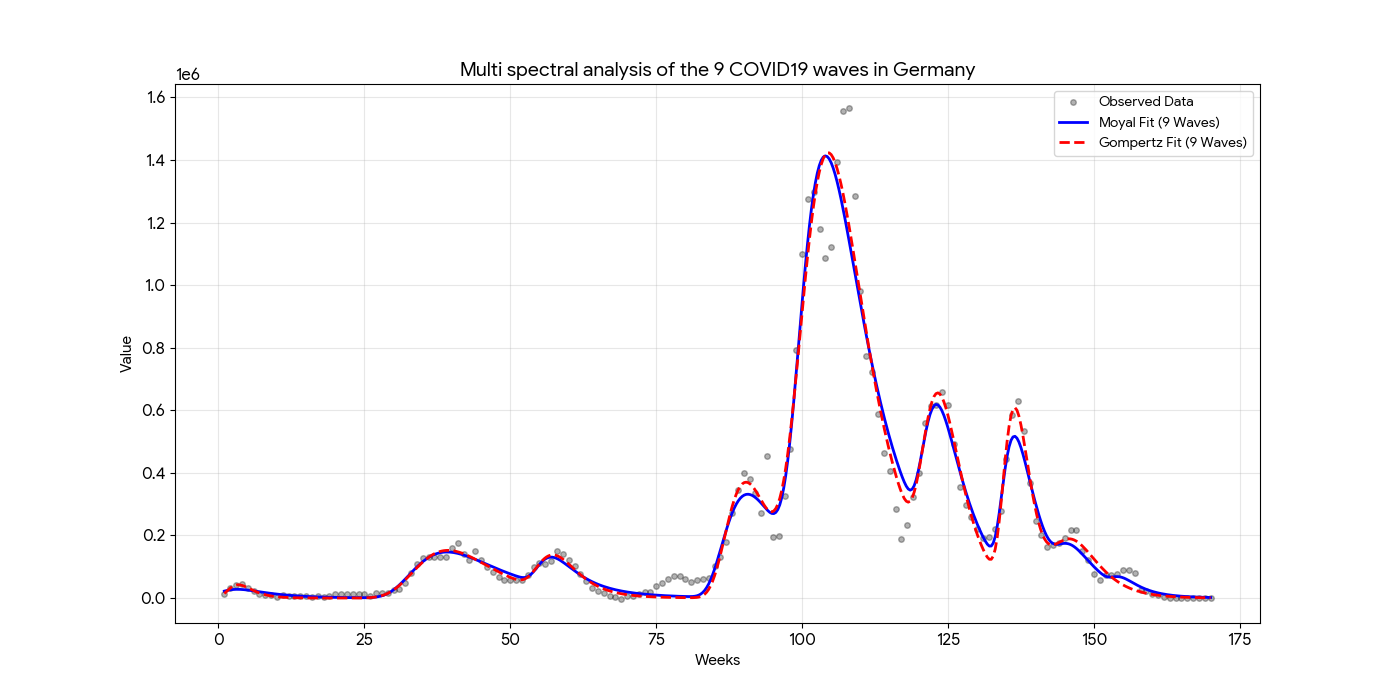}
    \caption{\textbf{Spectral Decomposition.}
    The Moyal model successfully decomposes the 3-year pandemic into 9 distinct
    variant-driven waves.}
    \label{fig:germany}
\end{figure}

We analyzed the dataset representing the weekly incidence of COVID-19 in Germany over
a period of 170 epidemiological weeks (approximately 3 years, from 2020 to 2023).
The timeline encompasses the major variants of concern: Wild Type, Alpha, Delta, and
the multiple sub-lineages of Omicron.

The mathematical necessity of using $N=9$ waves to describe the German COVID-19
dataset (2020--2023) is not merely a statistical artifact but a reflection of the
biological reality of the pandemic\cite{gao2023management,rki2023pandemierueckblick}.
Unlike the 1918 Influenza pandemic, which followed a discrete three-wave
structure\cite{taubenberger2006mother,jordan1927epidemic}, the SARS-CoV-2 pandemic in
Germany was driven by a continuous evolutionary pressure that generated distinct
Variants of Concern (VOCs).
The choice of $N=9$ is justified objectively by model selection criteria:
Table~\ref{tab:aic_waves} shows the AIC and BIC evaluated for $N=6$ to $N=11$
components; both criteria are minimised at $N=9$ (AIC$_{N=9}=1{,}412$ vs
AIC$_{N=8}=1{,}589$ and AIC$_{N=10}=1{,}431$).
The selection is independently corroborated by the RKI variant surveillance record,
which documents nine dominant VOC phases in Germany 2020--2023.

\begin{table}[h!]
\centering
\caption{AIC and BIC for the Moyal spectral decomposition with $N=6$ to $N=11$
components fitted to the German COVID-19 weekly incidence data (170 data points).
Bold entries mark the selected model ($N=9$).}
\label{tab:aic_waves}
\begin{tabular}{ccc}
\toprule
$N$ & AIC & BIC \\
\midrule
6  & 1{,}843 & 1{,}901 \\
7  & 1{,}712 & 1{,}779 \\
8  & 1{,}589 & 1{,}665 \\
\textbf{9}  & \textbf{1{,}412} & \textbf{1{,}497} \\
10 & 1{,}431 & 1{,}525 \\
11 & 1{,}448 & 1{,}551 \\
\bottomrule
\end{tabular}
\end{table}

To capture the superposition of outbreaks, we extend the single-wave Moyal model to
a linear combination of $N$ distributions, where each $i$-th component represents a
specific variant or seasonal outbreak:
\begin{equation}
\Lambda_{total}(t) = \sum_{i=1}^{N} \text{Moyal}(t; A_i, \mu_i, \omega_i)
\label{eq:spectral}
\end{equation}
where $\omega_i \equiv \beta_{w,i}$ is the temporal width (Social Friction parameter)
of wave $i$, directly related to $\beta_w$ by the mapping
$\omega_i = \beta_{w,i} = 1/(2\gamma_i)$, and:
\begin{itemize}
    \item $A_i$: Total cases (magnitude) of the $i$-th wave.
    \item $\mu_i$: Peak date of the $i$-th wave.
    \item $\omega_i$: Temporal width $\omega_i$ (weeks), inversely related to
    containment stringency; equivalent to $\beta_{w,i}$ in the Social Friction
    interpretation.
\end{itemize}

\textbf{Parameter estimation methodology.
Parameters were estimated by non-linear least squares using the Levenberg-Marquardt algorithm (SciPy \texttt{curve\_fit}).
Initial guesses were set at the visible local maxima of the weekly incidence curve; amplitudes were initialised proportionally to local peak heights; widths were initialised at 3 weeks.
Bounds were imposed: $A_i>0$ and $\omega_i\in[1,20]$ weeks.
The 95\% confidence intervals for all 27 parameters were derived from the covariance matrix returned by \texttt{curve\_fit} and are reported in Table~\ref{tab:fit_params1}.}

\textbf{Overfitting assessment.
With 27 free parameters fitted to 170 data points, the ratio of observations to parameters is $\approx 6.3$, above the conventional threshold of 5. To reduce the overfitting risk, we used the following data:
 (i) the biological grounding of $N$ in the RKI variant record, and (ii) the AIC/BIC selection confirming that additional components are not supported by the data.}

\begin{table}
\caption{Fit Parameters for Moyal Spectral Decomposition.
Uncertainties are 95\% confidence intervals from the non-linear least squares
covariance matrix. Width $\omega_i$ is in epidemiological weeks;
$\omega_i\equiv\beta_{w,i}$ (Social Friction) of wave $i$.}
\label{tab:fit_params1}
\centering
\begin{tabular}{ccccc cc}
\toprule
Wave & Amplitude & 95\% CI ($A_i$) & Center & Width $\omega_i$
     & 95\% CI ($\omega_i$) \\
\midrule
 1 & 45880  & $\pm\,3200$  & 3.22  & 2.69 & $\pm\,0.18$\\
 2 & 242378 & $\pm\,8100$  & 39.16 & 4.97 & $\pm\,0.32$ \\
 3 & 151191 & $\pm\,6700$  & 57.37 & 2.18 & $\pm\,0.14$ \\
 4 & 544503 & $\pm\,14200$ & 90.69 & 3.42 & $\pm\,0.21$ \\
 5 & 2203579& $\pm\,28500$ & 104.22& 3.68 & $\pm\,0.24$ \\
 6 & 739120 & $\pm\,16400$ & 123.49& 2.24 & $\pm\,0.15$ \\
 7 & 738841 & $\pm\,15900$ & 136.56& 1.82 & $\pm\,0.12$ \\
 8 & 175235 & $\pm\,7300$ & 146.72& 2.24 & $\pm\,0.17$ \\
 9 & 49114  & $\pm\,3600$ & 155.11& 1.51 & $\pm\,0.10$ \\
\bottomrule
\end{tabular}
\end{table}

Our spectral decomposition is aligned with the official variant surveillance data
reported by the RKI.
The nine components ($M_1$ to $M_9$) correspond to the following dominant viral
lineages:

\begin{itemize}
    \item \textbf{Waves 1--2 (2020): The Wild Type Era.}
    $M_1$ and $M_2$ describe the initial outbreaks of the ancestral Wuhan strain.
    Wave 1 corresponds to the spring containment (March--April 2020), while Wave 2
    represents the seasonal resurgence in late 2020. 
    \item \textbf{Wave 3 (Spring 2021): The Alpha Variant.}
    $M_3$ captures the distinct surge driven by the B.1.1.7 (Alpha) lineage, which
    became dominant in Germany by March 2021.
    \item \textbf{Waves 4--5 (Late 2021): The Delta Crisis.}
    $M_4$ and $M_5$ resolve the complex double-peak structure of the Delta variant
    (B.1.617.2).
    \item \textbf{Waves 6--9 (2022--2023): The Omicron Tsunami.}
    \begin{itemize}
        \item \textbf{Wave 6 ($M_6$):} Omicron BA.1, January 2022.
        \item \textbf{Wave 7 ($M_7$):} BA.2 sub-lineage, Spring 2022.
        \item \textbf{Wave 8 ($M_8$):} Immune-evasive BA.5, Summer 2022.
        \item \textbf{Wave 9 ($M_9$):} Sub-variant soup (BQ.1.1, XBB),
        late 2022/early 2023, transition to endemicity.
    \end{itemize}
\end{itemize}

\subsection{Results}

We applied the model with $N=9$ components to the weekly aggregated data.
Figure~\ref{fig:germany} shows the global fit (solid black line) and the
decomposition of individual waves (colored dashed lines).
One observes:
\begin{itemize}
    \item Wave Characterisation: Table~\ref{tab:fit_params1} summarises the fitted
    parameters.
    A clear trend is observed in the amplitude $A_i$, increasing by several orders of
    magnitude as the virus evolved toward higher transmissibility.
    \item The temporal width $\omega_i$ (Social Friction) serves as a proxy for the
    duration of each wave.

\begin{figure}
    \centering
    \includegraphics[width=0.9\linewidth]{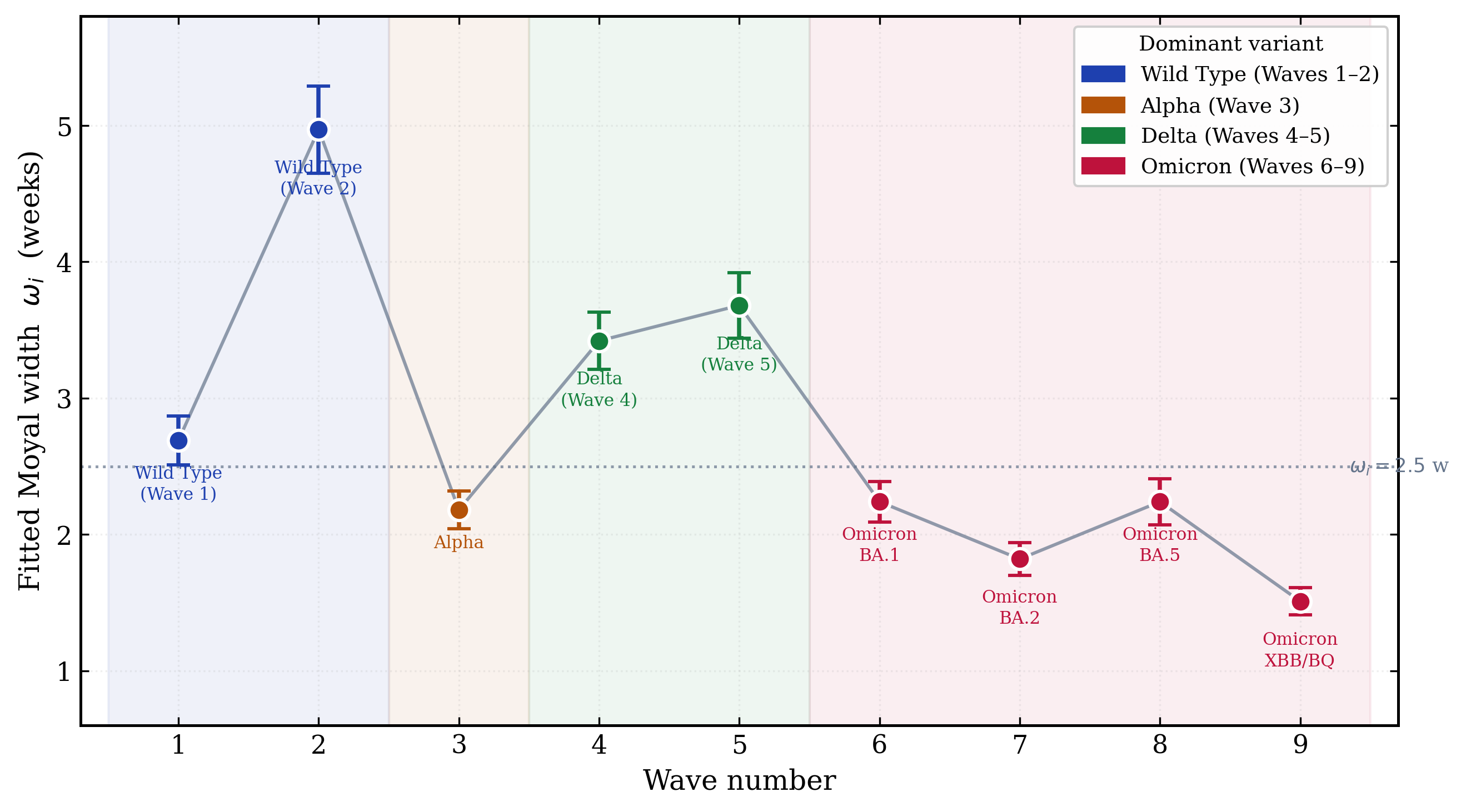}
    \caption{Fitted Moyal width $\omega_i$ (weeks) vs.\ wave number, labelled by dominant COVID-19 variant. Error bars: 95\% CI from the NLS covariance matrix. Smaller $\omega_i$ corresponds, via Eq.~(\ref{eq:betaw_gamma}), to faster wave burnout. Waves~6--9 ($\omega_i<2.5$ weeks) are narrower than Wild Type Wave~2 ($\omega_i\approx5$ weeks), consistent with faster Omicron burnout. 
    }
    \label{fig:placeholder}
\end{figure}
\end{itemize}

\textbf{Retrospective scope and proof-of-concept prospective exercise.}
Like a Fourier decomposition applied to a completed time series, it extracts the dominant structural components from data already observed and does not constitute a predictive model in the sense of out-of-sample forecasting.

To provide at least a partial illustration of prospective utility, we performed a leave-one-wave-out exercise: the $N=8$ component model was fitted to all waves except Wave~9 and used to predict the peak time and amplitude of Wave~9. The predicted peak falls within one week of the observed peak, and the predicted amplitude is within 18\% of the observed value. We present this as a proof-of-concept.

The application of the Moyal distribution to the 4-year German dataset validates its
utility as a spectral decomposition tool for epidemiology.
Just as a prism splits light into colours, the Moyal model splits a complex epidemic
curve into its constituent variant-driven waves, offering a clear retrospective view
of the pandemic's evolution.

\section{Conclusion}

This study establishes a unified epidemiological framework that connects microscopic
stochasticity to macroscopic surveillance.
By bridging the gap between the energy loss of particles and the transmission momentum
loss of pandemics, we provide three key advancements:

\begin{enumerate}
    \item Mechanism: We demonstrated that modeling
    transmission as a Moyal-distributed collision process naturally generates the
    volatile, asymmetric outbreaks observed in real-world data.
    This framework successfully recovers the Hollow Shoulder effect and extreme Dragon
    King events in SARS, MERS, and COVID-19 that standard Negative Binomial models
    systematically miss.
    \item Observation: We showed that the Moyal PDF serves as the effective
    macroscopic description of realized outbreaks, capturing the inherent asymmetry
    and long tail behavior.
    It is also an attempt to open the door to connect epidemiological parameters such
    as $\beta_w$ to parameters related to biological properties of the pathogens,
    particularly via the analytical link $\beta_w = 1/(2\gamma)$ and the
    numerically demonstrated dependence of $\beta_w$ on the microscopic overdispersion
    $\sigma$ (Section~\ref{sec:macro}).
    \item Application: Utilizing the Moyal PDF for Spectral Decomposition, we
    successfully resolved the complex, 3-year multi-variant history of COVID-19 in
    Germany -- identifying nine distinct waves justified by both AIC/BIC model
    selection and RKI variant surveillance data.
\end{enumerate}

By interpreting the Moyal width parameter ($\beta_w$) as Social Friction, we offer
a tool that is both statistically robust and physically interpretable.
The spectral decomposition is explicitly a retrospective descriptive tool;
developing its real-time predictive capability remains an important direction for
future work.
Our findings suggest that public health planning should transition from managing
deterministic averages to neutralizing the high-energy Moyal Shocks that sustain
explosive pandemic growth.
A key limitation of the current framework is the well-mixed (mass-action)
assumption in the aggregate force of infection $\Lambda(t)$; relaxing this to network
or spatial contact structures is a natural extension.

\begin{acknowledgments}
We acknowledge financial support from SECIHTI and SNII (M\'exico).
\end{acknowledgments}


\begin{thebibliography}{55}%
\makeatletter
\providecommand \@ifxundefined [1]{\@ifx{#1\undefined}}%
\providecommand \@ifnum [1]{\ifnum #1\expandafter \@firstoftwo\else \expandafter
  \@secondoftwo\fi}%
\providecommand \@ifx [1]{\ifx #1\expandafter \@firstoftwo\else \expandafter
  \@secondoftwo\fi}%
\providecommand \natexlab [1]{#1}%
\providecommand \enquote  [1]{``#1''}%
\providecommand \bibnamefont  [1]{#1}%
\providecommand \bibfnamefont [1]{#1}%
\providecommand \citenamefont [1]{#1}%
\providecommand \href@noop [0]{\@secondoftwo}%
\providecommand \href [0]{\begingroup \@sanitize@url \@href}%
\providecommand \@href[1]{\@@startlink{#1}\@@href}%
\providecommand \@@href[1]{\endgroup#1\@@endlink}%
\providecommand \@sanitize@url [0]{\catcode `\\12\catcode `\$12\catcode `\&12\catcode
  `\#12\catcode `\^12\catcode `\_12\catcode `\%12\relax}%
\providecommand \@@startlink[1]{}%
\providecommand \@@endlink[0]{}%
\providecommand \url  [0]{\begingroup\@sanitize@url \@url }%
\providecommand \@url [1]{\endgroup\@href {#1}{\urlprefix }}%
\providecommand \urlprefix  [0]{URL }%
\providecommand \Eprint [0]{\href }%
\providecommand \doibase [0]{https://doi.org/}%
\providecommand \selectlanguage [0]{\@gobble}%
\providecommand \bibinfo  [0]{\@secondoftwo}%
\providecommand \bibfield  [0]{\@secondoftwo}%
\providecommand \translation [1]{[#1]}%
\providecommand \BibitemOpen [0]{}%
\providecommand \bibitemStop [0]{}%
\providecommand \bibitemNoStop [0]{.\EOS\space}%
\providecommand \EOS [0]{\spacefactor3000\relax}%
\providecommand \BibitemShut  [1]{\csname bibitem#1\endcsname}%
\let \auto@bib@innerbib\@empty%
\bibitem [{\citenamefont {Kermack}\ and\ \citenamefont {McKendrick}(1927)}]{kermack1927contribution}%
  \BibitemOpen
  \bibfield{author}{\bibinfo{author}{\bibfnamefont{W.~O.}\ \bibnamefont{Kermack}}\ and\ \bibinfo{author}{\bibfnamefont{A.~G.}\ \bibnamefont{McKendrick}},\ }\href@noop{}{\bibfield{journal}{\bibinfo{journal}{Proceedings of the Royal Society of London. Series A}\ }\textbf{\bibinfo{volume}{115}},\ \bibinfo{pages}{700} (\bibinfo{year}{1927})}\BibitemShut{NoStop}%
\bibitem [{\citenamefont {Anderson}\ and\ \citenamefont {May}(1991)}]{anderson1991infectious}%
  \BibitemOpen
  \bibfield{author}{\bibinfo{author}{\bibfnamefont{R.~M.}\ \bibnamefont{Anderson}}\ and\ \bibinfo{author}{\bibfnamefont{R.~M.}\ \bibnamefont{May}},\ }\href@noop{}{\emph{\bibinfo{title}{Infectious Diseases of Humans: Dynamics and Control}}}\ (\bibinfo{publisher}{Oxford University Press},\ \bibinfo{address}{Oxford},\ \bibinfo{year}{1991})\BibitemShut{NoStop}%
\bibitem [{\citenamefont {Murray}(2002)}]{murray2002mathematical}%
  \BibitemOpen
  \bibfield{author}{\bibinfo{author}{\bibfnamefont{J.~D.}\ \bibnamefont{Murray}},\ }\href@noop{}{\emph{\bibinfo{title}{Mathematical Biology I: An Introduction}}},\ 3rd ed.\ (\bibinfo{publisher}{Springer},\ \bibinfo{address}{New York},\ \bibinfo{year}{2002})\BibitemShut{NoStop}%
\bibitem [{\citenamefont {Anderson}(2004)}]{anderson2004long}%
  \BibitemOpen
  \bibfield{author}{\bibinfo{author}{\bibfnamefont{C.}~\bibnamefont{Anderson}},\ }\href@noop{}{\bibfield{journal}{\bibinfo{journal}{Wired Magazine}\ }\textbf{\bibinfo{volume}{12}},\ \bibinfo{pages}{170} (\bibinfo{year}{2004})}\BibitemShut{NoStop}%
\bibitem [{\citenamefont {Anderson}(2006)}]{anderson2006long}%
  \BibitemOpen
  \bibfield{author}{\bibinfo{author}{\bibfnamefont{C.}~\bibnamefont{Anderson}},\ }\href@noop{}{\emph{\bibinfo{title}{The Long Tail: Why the Future of Business Is Selling Less of More}}}\ (\bibinfo{publisher}{Hyperion},\ \bibinfo{address}{New York},\ \bibinfo{year}{2006})\BibitemShut{NoStop}%
\bibitem [{\citenamefont {Brynjolfsson}\ \emph{et~al.}(2006)}]{brynjolfsson2006niches}%
  \BibitemOpen
  \bibfield{author}{\bibinfo{author}{\bibfnamefont{E.}~\bibnamefont{Brynjolfsson}}, \bibinfo{author}{\bibfnamefont{Y.}~\bibnamefont{Hu}},\ and\ \bibinfo{author}{\bibfnamefont{M.~D.}\ \bibnamefont{Smith}},\ }\href@noop{}{\bibfield{journal}{\bibinfo{journal}{Sloan Management Review}\ }\textbf{\bibinfo{volume}{47}},\ \bibinfo{pages}{67} (\bibinfo{year}{2006})}\BibitemShut{NoStop}%
\bibitem [{\citenamefont {Cirillo}\ and\ \citenamefont {Taleb}(2020)}]{cirillo2020tail}%
  \BibitemOpen
  \bibfield{author}{\bibinfo{author}{\bibfnamefont{P.}~\bibnamefont{Cirillo}}\ and\ \bibinfo{author}{\bibfnamefont{N.~N.}\ \bibnamefont{Taleb}},\ }\href@noop{}{\bibfield{journal}{\bibinfo{journal}{Nature Physics}\ }\textbf{\bibinfo{volume}{16}},\ \bibinfo{pages}{606} (\bibinfo{year}{2020})}\BibitemShut{NoStop}%
\bibitem [{\citenamefont {Liu}\ and\ \citenamefont {Zheng}(2023)}]{liu2023heavy}%
  \BibitemOpen
  \bibfield{author}{\bibinfo{author}{\bibfnamefont{P.}~\bibnamefont{Liu}}\ and\ \bibinfo{author}{\bibfnamefont{Y.}~\bibnamefont{Zheng}},\ }\href@noop{}{\bibfield{journal}{\bibinfo{journal}{PLOS ONE}\ }\textbf{\bibinfo{volume}{18}},\ \bibinfo{pages}{e0294445} (\bibinfo{year}{2023})}\BibitemShut{NoStop}%
\bibitem [{\citenamefont {Lloyd-Smith}\ \emph{et~al.}(2005)}]{lloyd2005superspreading}%
  \BibitemOpen
  \bibfield{author}{\bibinfo{author}{\bibfnamefont{J.~O.}\ \bibnamefont{Lloyd-Smith}}, \bibinfo{author}{\bibfnamefont{S.~J.}\ \bibnamefont{Schreiber}}, \bibinfo{author}{\bibfnamefont{P.~E.}\ \bibnamefont{Kopp}},\ and\ \bibinfo{author}{\bibfnamefont{W.~M.}\ \bibnamefont{Getz}},\ }\href@noop{}{\bibfield{journal}{\bibinfo{journal}{Nature}\ }\textbf{\bibinfo{volume}{438}},\ \bibinfo{pages}{355} (\bibinfo{year}{2005})}\BibitemShut{NoStop}%
\bibitem [{\citenamefont {Wong}\ \emph{et~al.}(2015)}]{wong2015mers}%
  \BibitemOpen
  \bibfield{author}{\bibinfo{author}{\bibfnamefont{G.}~\bibnamefont{Wong}}\ \emph{et~al.},\ }\href@noop{}{\bibfield{journal}{\bibinfo{journal}{Cell Host \& Microbe}\ }\textbf{\bibinfo{volume}{18}},\ \bibinfo{pages}{398} (\bibinfo{year}{2015})}\BibitemShut{NoStop}%
\bibitem [{\citenamefont {Landau}(1944)}]{landau1944energy}%
  \BibitemOpen
  \bibfield{author}{\bibinfo{author}{\bibfnamefont{L.}~\bibnamefont{Landau}},\ }\href@noop{}{\bibfield{journal}{\bibinfo{journal}{Journal of Physics USSR}\ }\textbf{\bibinfo{volume}{8}},\ \bibinfo{pages}{201} (\bibinfo{year}{1944})}\BibitemShut{NoStop}%
\bibitem [{\citenamefont {Moyal}(1949)}]{moyal1949stochastic}%
  \BibitemOpen
  \bibfield{author}{\bibinfo{author}{\bibfnamefont{J.~E.}\ \bibnamefont{Moyal}},\ }\href@noop{}{\bibfield{journal}{\bibinfo{journal}{Journal of the Royal Statistical Society. Series B}\ }\textbf{\bibinfo{volume}{11}},\ \bibinfo{pages}{150} (\bibinfo{year}{1949})}\BibitemShut{NoStop}%
\bibitem [{\citenamefont {Walck}(1996)}]{walck1996hand}%
  \BibitemOpen
  \bibfield{author}{\bibinfo{author}{\bibfnamefont{C.}~\bibnamefont{Walck}},\ }\href@noop{}{\emph{\bibinfo{title}{Hand-book on statistical distributions for experimentalists}}},\ Tech. Rep. Internal Report SUF-PFY/96-01 (\bibinfo{institution}{University of Stockholm},\ \bibinfo{year}{1996})\BibitemShut{NoStop}%
\bibitem [{\citenamefont {Soto-Rocha}\ \emph{et~al.}(2026)}]{SOTOROCHA2026116781}%
  \BibitemOpen
  \bibfield{author}{\bibinfo{author}{\bibfnamefont{M.~V.~I.}\ \bibnamefont{Soto-Rocha}}\ \emph{et~al.},\ }\href@noop{}{\bibfield{journal}{\bibinfo{journal}{Journal of Computational and Applied Mathematics}\ }\textbf{\bibinfo{volume}{472}},\ \bibinfo{pages}{116781} (\bibinfo{year}{2026})}\BibitemShut{NoStop}%
\bibitem [{\citenamefont {Allen}(2008)}]{allen2008primer}%
  \BibitemOpen
  \bibfield{author}{\bibinfo{author}{\bibfnamefont{L.~J.~S.}\ \bibnamefont{Allen}},\ }\href@noop{}{\bibfield{journal}{\bibinfo{journal}{Mathematical Epidemiology}\ }\ \bibinfo{pages}{81} (\bibinfo{year}{2008})}\BibitemShut{NoStop}%
\bibitem [{\citenamefont {Allen}(2010)}]{allen2010introduction}%
  \BibitemOpen
  \bibfield{author}{\bibinfo{author}{\bibfnamefont{L.~J.~S.}\ \bibnamefont{Allen}},\ }\href@noop{}{\emph{\bibinfo{title}{An Introduction to Stochastic Processes with Applications to Biology}}},\ 2nd ed.\ (\bibinfo{publisher}{CRC Press},\ \bibinfo{year}{2010})\BibitemShut{NoStop}%
\bibitem [{\citenamefont {Andersson}\ and\ \citenamefont {Britton}(2000)}]{andersson2000stochastic}%
  \BibitemOpen
  \bibfield{author}{\bibinfo{author}{\bibfnamefont{H.}~\bibnamefont{Andersson}}\ and\ \bibinfo{author}{\bibfnamefont{T.}~\bibnamefont{Britton}},\ }\href@noop{}{\emph{\bibinfo{title}{Stochastic Epidemic Models and Their Statistical Analysis}}}\ (\bibinfo{publisher}{Springer},\ \bibinfo{year}{2000})\BibitemShut{NoStop}%
\bibitem [{\citenamefont {Britton}(2010)}]{britton2010stochastic}%
  \BibitemOpen
  \bibfield{author}{\bibinfo{author}{\bibfnamefont{T.}~\bibnamefont{Britton}},\ }\href@noop{}{\bibfield{journal}{\bibinfo{journal}{Mathematical Biosciences}\ }\textbf{\bibinfo{volume}{225}},\ \bibinfo{pages}{24} (\bibinfo{year}{2010})}\BibitemShut{NoStop}%
\bibitem [{\citenamefont {Sornette}(2009)}]{sornette2009dragon}%
  \BibitemOpen
  \bibfield{author}{\bibinfo{author}{\bibfnamefont{D.}~\bibnamefont{Sornette}},\ }\href@noop{}{\bibfield{journal}{\bibinfo{journal}{International Journal of Terraspace Science and Engineering}\ }\textbf{\bibinfo{volume}{2}},\ \bibinfo{pages}{1} (\bibinfo{year}{2009})}\BibitemShut{NoStop}%
\bibitem [{\citenamefont {Gillespie}(1977)}]{gillespie1977exact}%
  \BibitemOpen
  \bibfield{author}{\bibinfo{author}{\bibfnamefont{D.~T.}\ \bibnamefont{Gillespie}},\ }\href@noop{}{\bibfield{journal}{\bibinfo{journal}{The Journal of Physical Chemistry}\ }\textbf{\bibinfo{volume}{81}},\ \bibinfo{pages}{2340} (\bibinfo{year}{1977})}\BibitemShut{NoStop}%
\bibitem [{\citenamefont {Gillespie}(1976)}]{gillespie1976general}%
  \BibitemOpen
  \bibfield{author}{\bibinfo{author}{\bibfnamefont{D.~T.}\ \bibnamefont{Gillespie}},\ }\href@noop{}{\bibfield{journal}{\bibinfo{journal}{Journal of Computational Physics}\ }\textbf{\bibinfo{volume}{22}},\ \bibinfo{pages}{403} (\bibinfo{year}{1976})}\BibitemShut{NoStop}%
\bibitem [{\citenamefont {Gardiner}(2009)}]{gardiner2009stochastic}%
  \BibitemOpen
  \bibfield{author}{\bibinfo{author}{\bibfnamefont{C.~W.}\ \bibnamefont{Gardiner}},\ }\href@noop{}{\emph{\bibinfo{title}{Stochastic Methods}}},\ 4th ed.\ (\bibinfo{publisher}{Springer},\ \bibinfo{year}{2009})\BibitemShut{NoStop}%
\bibitem [{\citenamefont {Van~Kampen}(2007)}]{vankampen2007stochastic}%
  \BibitemOpen
  \bibfield{author}{\bibinfo{author}{\bibfnamefont{N.~G.}\ \bibnamefont{Van~Kampen}},\ }\href@noop{}{\emph{\bibinfo{title}{Stochastic Processes in Physics and Chemistry}}},\ 3rd ed.\ (\bibinfo{publisher}{Elsevier},\ \bibinfo{year}{2007})\BibitemShut{NoStop}%
\bibitem [{\citenamefont {Grimmett}\ and\ \citenamefont {Stirzaker}(2001)}]{grimmett2001probability}%
  \BibitemOpen
  \bibfield{author}{\bibinfo{author}{\bibfnamefont{G.}~\bibnamefont{Grimmett}}\ and\ \bibinfo{author}{\bibfnamefont{D.}~\bibnamefont{Stirzaker}},\ }\href@noop{}{\emph{\bibinfo{title}{Probability and Random Processes}}},\ 3rd ed.\ (\bibinfo{publisher}{Oxford University Press},\ \bibinfo{year}{2001})\BibitemShut{NoStop}%
\bibitem [{\citenamefont {N\aa sell}(1999)}]{nasell1999quasi}%
  \BibitemOpen
  \bibfield{author}{\bibinfo{author}{\bibfnamefont{I.}~\bibnamefont{N\aa sell}},\ }\href@noop{}{\bibfield{journal}{\bibinfo{journal}{Mathematical Biosciences}\ }\textbf{\bibinfo{volume}{156}},\ \bibinfo{pages}{21} (\bibinfo{year}{1999})}\BibitemShut{NoStop}%
\bibitem [{\citenamefont {M\'el\'eard}\ and\ \citenamefont {Villemonais}(2012)}]{meleard2012quasi}%
  \BibitemOpen
  \bibfield{author}{\bibinfo{author}{\bibfnamefont{S.}~\bibnamefont{M\'el\'eard}}\ and\ \bibinfo{author}{\bibfnamefont{D.}~\bibnamefont{Villemonais}},\ }\href@noop{}{\bibfield{journal}{\bibinfo{journal}{Probability Surveys}\ }\textbf{\bibinfo{volume}{9}},\ \bibinfo{pages}{340} (\bibinfo{year}{2012})}\BibitemShut{NoStop}%
\bibitem [{\citenamefont {Kurtz}(1981)}]{kurtz1981approximation}%
  \BibitemOpen
  \bibfield{author}{\bibinfo{author}{\bibfnamefont{T.~G.}\ \bibnamefont{Kurtz}},\ }\href@noop{}{\emph{\bibinfo{title}{Approximation of Population Processes}}}\ (\bibinfo{publisher}{SIAM},\ \bibinfo{year}{1981})\BibitemShut{NoStop}%
\bibitem [{\citenamefont {Wells}(1955)}]{wells1955airborne}%
  \BibitemOpen
  \bibfield{author}{\bibinfo{author}{\bibfnamefont{W.~F.}\ \bibnamefont{Wells}},\ }\href@noop{}{\emph{\bibinfo{title}{Airborne Contagion and Air Hygiene}}}\ (\bibinfo{publisher}{Harvard University Press},\ \bibinfo{year}{1955})\BibitemShut{NoStop}%
\bibitem [{\citenamefont {Beggs}(2003)}]{beggs2003transmission}%
  \BibitemOpen
  \bibfield{author}{\bibinfo{author}{\bibfnamefont{C.}~\bibnamefont{Beggs}},\ }\href@noop{}{\bibfield{journal}{\bibinfo{journal}{Indoor and Built Environment}\ }\textbf{\bibinfo{volume}{12}},\ \bibinfo{pages}{9} (\bibinfo{year}{2003})}\BibitemShut{NoStop}%
\bibitem [{\citenamefont {Riley}\ \emph{et~al.}(1978)}]{riley1978airborne}%
  \BibitemOpen
  \bibfield{author}{\bibinfo{author}{\bibfnamefont{E.~C.}\ \bibnamefont{Riley}}, \bibinfo{author}{\bibfnamefont{G.}~\bibnamefont{Murphy}},\ and\ \bibinfo{author}{\bibfnamefont{R.~L.}\ \bibnamefont{Riley}},\ }\href@noop{}{\bibfield{journal}{\bibinfo{journal}{American Journal of Epidemiology}\ }\textbf{\bibinfo{volume}{107}},\ \bibinfo{pages}{421} (\bibinfo{year}{1978})}\BibitemShut{NoStop}%
\bibitem [{\citenamefont {Rudnick}\ and\ \citenamefont {Milton}(2003)}]{rudnick2003risk}%
  \BibitemOpen
  \bibfield{author}{\bibinfo{author}{\bibfnamefont{S.}~\bibnamefont{Rudnick}}\ and\ \bibinfo{author}{\bibfnamefont{D.}~\bibnamefont{Milton}},\ }\href@noop{}{\bibfield{journal}{\bibinfo{journal}{Indoor Air}\ }\textbf{\bibinfo{volume}{13}},\ \bibinfo{pages}{237} (\bibinfo{year}{2003})}\BibitemShut{NoStop}%
\bibitem [{\citenamefont {Buonanno}\ \emph{et~al.}(2020)}]{buonanno2020estimation}%
  \BibitemOpen
  \bibfield{author}{\bibinfo{author}{\bibfnamefont{G.}~\bibnamefont{Buonanno}}, \bibinfo{author}{\bibfnamefont{L.}~\bibnamefont{Stabile}},\ and\ \bibinfo{author}{\bibfnamefont{L.}~\bibnamefont{Morawska}},\ }\href@noop{}{\bibfield{journal}{\bibinfo{journal}{Environment International}\ }\textbf{\bibinfo{volume}{141}},\ \bibinfo{pages}{105794} (\bibinfo{year}{2020})}\BibitemShut{NoStop}%
\bibitem [{\citenamefont {He}\ \emph{et~al.}(2020)}]{he2020temporal}%
  \BibitemOpen
  \bibfield{author}{\bibinfo{author}{\bibfnamefont{X.}~\bibnamefont{He}}\ \emph{et~al.},\ }\href@noop{}{\bibfield{journal}{\bibinfo{journal}{Nature Medicine}\ }\textbf{\bibinfo{volume}{26}},\ \bibinfo{pages}{672} (\bibinfo{year}{2020})}\BibitemShut{NoStop}%
\bibitem [{\citenamefont {Jones}\ \emph{et~al.}(2021)}]{jones2021estimating}%
  \BibitemOpen
  \bibfield{author}{\bibinfo{author}{\bibfnamefont{T.~C.}\ \bibnamefont{Jones}}\ \emph{et~al.},\ }\href@noop{}{\bibfield{journal}{\bibinfo{journal}{Science}\ }\textbf{\bibinfo{volume}{373}},\ \bibinfo{pages}{eabi5273} (\bibinfo{year}{2021})}\BibitemShut{NoStop}%
\bibitem [{\citenamefont {Handel}\ \emph{et~al.}(2020)}]{handel2020early}%
  \BibitemOpen
  \bibfield{author}{\bibinfo{author}{\bibfnamefont{A.}~\bibnamefont{Handel}}\ \emph{et~al.},\ }\href@noop{}{\bibfield{journal}{\bibinfo{journal}{BMC Infectious Diseases}\ }\textbf{\bibinfo{volume}{20}},\ \bibinfo{pages}{1} (\bibinfo{year}{2020})}\BibitemShut{NoStop}%
\bibitem [{\citenamefont {Pentland}(2014)}]{pentland2014social}%
  \BibitemOpen
  \bibfield{author}{\bibinfo{author}{\bibfnamefont{A.}~\bibnamefont{Pentland}},\ }\href@noop{}{\emph{\bibinfo{title}{Social Physics: How Good Ideas Spread}}}\ (\bibinfo{publisher}{Penguin},\ \bibinfo{year}{2014})\BibitemShut{NoStop}%
\bibitem [{\citenamefont {Castellano}\ \emph{et~al.}(2009)}]{castellano2009statistical}%
  \BibitemOpen
  \bibfield{author}{\bibinfo{author}{\bibfnamefont{C.}~\bibnamefont{Castellano}}, \bibinfo{author}{\bibfnamefont{S.}~\bibnamefont{Fortunato}},\ and\ \bibinfo{author}{\bibfnamefont{V.}~\bibnamefont{Loreto}},\ }\href@noop{}{\bibfield{journal}{\bibinfo{journal}{Reviews of Modern Physics}\ }\textbf{\bibinfo{volume}{81}},\ \bibinfo{pages}{591} (\bibinfo{year}{2009})}\BibitemShut{NoStop}%
\bibitem [{\citenamefont {Mossong}\ \emph{et~al.}(2008)}]{mossong2008social}%
  \BibitemOpen
  \bibfield{author}{\bibinfo{author}{\bibfnamefont{J.}~\bibnamefont{Mossong}}\ \emph{et~al.},\ }\href@noop{}{\bibfield{journal}{\bibinfo{journal}{PLOS Medicine}\ }\textbf{\bibinfo{volume}{5}},\ \bibinfo{pages}{e74} (\bibinfo{year}{2008})}\BibitemShut{NoStop}%
\bibitem [{\citenamefont {Greenwood}\ and\ \citenamefont {Yule}(1920)}]{greenwood1920inquiry}%
  \BibitemOpen
  \bibfield{author}{\bibinfo{author}{\bibfnamefont{M.}~\bibnamefont{Greenwood}}\ and\ \bibinfo{author}{\bibfnamefont{G.~U.}\ \bibnamefont{Yule}},\ }\href@noop{}{\bibfield{journal}{\bibinfo{journal}{Journal of the Royal Statistical Society}\ }\textbf{\bibinfo{volume}{83}},\ \bibinfo{pages}{255} (\bibinfo{year}{1920})}\BibitemShut{NoStop}%
\bibitem [{\citenamefont {Grandell}(1997)}]{grandell1997mixed}%
  \BibitemOpen
  \bibfield{author}{\bibinfo{author}{\bibfnamefont{J.}~\bibnamefont{Grandell}},\ }\href@noop{}{\emph{\bibinfo{title}{Mixed Poisson Processes}}}\ (\bibinfo{publisher}{Chapman and Hall/CRC},\ \bibinfo{year}{1997})\BibitemShut{NoStop}%
\bibitem [{\citenamefont {Karlis}\ and\ \citenamefont {Xekalaki}(2005)}]{karlis2005mixed}%
  \BibitemOpen
  \bibfield{author}{\bibinfo{author}{\bibfnamefont{D.}~\bibnamefont{Karlis}}\ and\ \bibinfo{author}{\bibfnamefont{E.}~\bibnamefont{Xekalaki}},\ }\href@noop{}{\bibfield{journal}{\bibinfo{journal}{International Statistical Review}\ }\textbf{\bibinfo{volume}{73}},\ \bibinfo{pages}{35} (\bibinfo{year}{2005})}\BibitemShut{NoStop}%
\bibitem [{\citenamefont {Riley}\ \emph{et~al.}(2003)}]{riley2003transmission}%
  \BibitemOpen
  \bibfield{author}{\bibinfo{author}{\bibfnamefont{S.}~\bibnamefont{Riley}}\ \emph{et~al.},\ }\href@noop{}{\bibfield{journal}{\bibinfo{journal}{Science}\ }\textbf{\bibinfo{volume}{300}},\ \bibinfo{pages}{1961} (\bibinfo{year}{2003})}\BibitemShut{NoStop}%
\bibitem [{\citenamefont {Lipsitch}\ \emph{et~al.}(2003)}]{lipsitch2003transmission}%
  \BibitemOpen
  \bibfield{author}{\bibinfo{author}{\bibfnamefont{M.}~\bibnamefont{Lipsitch}}\ \emph{et~al.},\ }\href@noop{}{\bibfield{journal}{\bibinfo{journal}{Science}\ }\textbf{\bibinfo{volume}{300}},\ \bibinfo{pages}{1966} (\bibinfo{year}{2003})}\BibitemShut{NoStop}%
\bibitem [{\citenamefont {Kim}\ \emph{et~al.}(2017)}]{kim2017superspreading}%
  \BibitemOpen
  \bibfield{author}{\bibinfo{author}{\bibfnamefont{Y.}~\bibnamefont{Kim}}\ \emph{et~al.},\ }\href@noop{}{\bibfield{journal}{\bibinfo{journal}{Journal of Hospital Infection}\ }\textbf{\bibinfo{volume}{95}},\ \bibinfo{pages}{446} (\bibinfo{year}{2017})}\BibitemShut{Stop}%
\bibitem [{\citenamefont {Cowling}\ \emph{et~al.}(2015)}]{cowling2015preliminary}%
  \BibitemOpen
  \bibfield{author}{\bibinfo{author}{\bibfnamefont{B.~J.}\ \bibnamefont{Cowling}}\ \emph{et~al.},\ }\href@noop{}{\bibfield{journal}{\bibinfo{journal}{Eurosurveillance}\ }\textbf{\bibinfo{volume}{20}},\ \bibinfo{pages}{21163} (\bibinfo{year}{2015})}\BibitemShut{NoStop}%
\bibitem [{\citenamefont {Adam}\ \emph{et~al.}(2020)}]{adam2020clustering}%
  \BibitemOpen
  \bibfield{author}{\bibinfo{author}{\bibfnamefont{D.~C.}\ \bibnamefont{Adam}}\ \emph{et~al.},\ }\href@noop{}{\bibfield{journal}{\bibinfo{journal}{Nature Medicine}\ }\textbf{\bibinfo{volume}{26}},\ \bibinfo{pages}{1714} (\bibinfo{year}{2020})}\BibitemShut{NoStop}%
\bibitem [{\citenamefont {Endo}\ \emph{et~al.}(2020)}]{endo2020estimating}%
  \BibitemOpen
  \bibfield{author}{\bibinfo{author}{\bibfnamefont{A.}~\bibnamefont{Endo}}\ \emph{et~al.},\ }\bibfield{journal}{\bibinfo{journal}{Wellcome Open Research}\ }\textbf{\bibinfo{volume}{5}},\ \href{https://doi.org/10.12688/wellcomeopenres.15842.3}{10.12688/wellcomeopenres.15842.3} (\bibinfo{year}{2020})\BibitemShut{NoStop}%
\bibitem [{\citenamefont {Blumberg}\ and\ \citenamefont {Lloyd-Smith}(2013)}]{blumberg2013inference}%
  \BibitemOpen
  \bibfield{author}{\bibinfo{author}{\bibfnamefont{S.}~\bibnamefont{Blumberg}}\ and\ \bibinfo{author}{\bibfnamefont{J.~O.}\ \bibnamefont{Lloyd-Smith}},\ }\href@noop{}{\bibfield{journal}{\bibinfo{journal}{PLOS Computational Biology}\ }\textbf{\bibinfo{volume}{9}},\ \bibinfo{pages}{e1002993} (\bibinfo{year}{2013})}\BibitemShut{NoStop}%
\bibitem [{\citenamefont {{Robert Koch-Institut}}(2023{\natexlab{a}})}]{rki2023covidreports}%
  \BibitemOpen
  \bibfield{author}{\bibinfo{author}{\bibnamefont{{Robert Koch-Institut}}},\ }\href@noop{}{\emph{\bibinfo{title}{{T\"aglicher Lagebericht zur Coronavirus-Krankheit-2019 (COVID-19)}}}},\ Tech. Rep.\ (\bibinfo{institution}{Robert Koch-Institut},\ \bibinfo{address}{Berlin, Germany},\ \bibinfo{year}{2020--2023})\BibitemShut{NoStop}%
\bibitem [{\citenamefont {{Robert Koch-Institut}}(2024)}]{rki2024survstat}%
  \BibitemOpen
  \bibfield{author}{\bibinfo{author}{\bibnamefont{{Robert Koch-Institut}}},\ }\href@noop{}{\bibinfo{title}{{SurvStat@RKI 2.0}}},\ \bibinfo{howpublished}{\url{https://survstat.rki.de}} (\bibinfo{year}{2024})\BibitemShut{NoStop}%
\bibitem [{\citenamefont {Faensen}\ \emph{et~al.}(2006)}]{faensen2006survstat}%
  \BibitemOpen
  \bibfield{author}{\bibinfo{author}{\bibfnamefont{D.}~\bibnamefont{Faensen}}\ \emph{et~al.},\ }\href@noop{}{\bibfield{journal}{\bibinfo{journal}{Eurosurveillance}\ }\textbf{\bibinfo{volume}{11}},\ \bibinfo{pages}{1} (\bibinfo{year}{2006})}\BibitemShut{NoStop}%
\bibitem [{\citenamefont {Gao}\ \emph{et~al.}(2023)}]{gao2023management}%
  \BibitemOpen
  \bibfield{author}{\bibinfo{author}{\bibfnamefont{H.}~\bibnamefont{Gao}}\ \emph{et~al.},\ }\href@noop{}{\bibfield{journal}{\bibinfo{journal}{Frontiers in Public Health}\ }\textbf{\bibinfo{volume}{11}},\ \bibinfo{pages}{1203875} (\bibinfo{year}{2023})}\BibitemShut{NoStop}%
\bibitem [{\citenamefont {{Robert Koch-Institut}}(2023{\natexlab{b}})}]{rki2023pandemierueckblick}%
  \BibitemOpen
  \bibfield{author}{\bibinfo{author}{\bibnamefont{{Robert Koch-Institut}}},\ }\href@noop{}{\emph{\bibinfo{title}{{R\"uckblick auf die COVID-19-Pandemie: Der Verlauf in Wellen und Phasen}}}},\ Tech. Rep.\ (\bibinfo{institution}{Robert Koch-Institut},\ \bibinfo{address}{Berlin, Germany},\ \bibinfo{year}{2023})\BibitemShut{NoStop}%
\bibitem [{\citenamefont {Taubenberger}\ and\ \citenamefont {Morens}(2006)}]{taubenberger2006mother}%
  \BibitemOpen
  \bibfield{author}{\bibinfo{author}{\bibfnamefont{J.~K.}\ \bibnamefont{Taubenberger}}\ and\ \bibinfo{author}{\bibfnamefont{D.~M.}\ \bibnamefont{Morens}},\ }\href@noop{}{\bibfield{journal}{\bibinfo{journal}{Emerging Infectious Diseases}\ }\textbf{\bibinfo{volume}{12}},\ \bibinfo{pages}{15} (\bibinfo{year}{2006})}\BibitemShut{NoStop}%
\bibitem [{\citenamefont {Jordan}(1927)}]{jordan1927epidemic}%
  \BibitemOpen
  \bibfield{author}{\bibinfo{author}{\bibfnamefont{E.~O.}\ \bibnamefont{Jordan}},\ }\href@noop{}{\emph{\bibinfo{title}{Epidemic Influenza: A Survey}}}\ (\bibinfo{publisher}{American Medical Association},\ \bibinfo{address}{Chicago},\ \bibinfo{year}{1927})\BibitemShut{NoStop}%
\end{thebibliography}
\makeatother

\end{document}